\documentclass[aps,twocolumn,groupedaddress,superscriptaddress,amsmath,amssymb,prb,longbibliography]{revtex4-2}
\usepackage{mathtools,amsmath,amsxtra}
\usepackage{amssymb}
\usepackage{physics}
\usepackage[english]{babel}
\usepackage{dsfont}
\usepackage{graphicx}
\usepackage{dcolumn}
\usepackage{bm}
\usepackage{bbold}
\usepackage[usenames,dvipsnames]{color}
\usepackage[colorlinks,citecolor=Blue,linkcolor=Red, urlcolor=Blue]{hyperref}
\usepackage{tikz}
\usetikzlibrary{decorations.pathmorphing}
\usetikzlibrary{arrows.meta}
\usepackage{braket}
\usepackage[normalem]{ulem}
\usepackage{lipsum}
\usepackage{caption}
\DeclareCaptionType{equ}[][]

\begin{document}

\title{Quantum Information Processing with Molecular Nanomagnets: an introduction} 

\author{Alessandro Chiesa}
\email{alessandro.chiesa@unipr.it}
\affiliation{Università di Parma, Dipartimento di Scienze Matematiche, Fisiche e Informatiche, I-43124, Parma, Italy}
\affiliation{UdR Parma, INSTM, I-43124 Parma, Italy}
\affiliation{Gruppo Collegato di Parma, INFN-Sezione Milano-Bicocca, I-43124 Parma, Italy}

\author{Emilio Macaluso}
\affiliation{Università di Parma, Dipartimento di Scienze Matematiche, Fisiche e Informatiche, I-43124, Parma, Italy}

\author{Stefano Carretta}
\email{stefano.carretta@unipr.it}
\affiliation{Università di Parma, Dipartimento di Scienze Matematiche, Fisiche e Informatiche, I-43124, Parma, Italy}
\affiliation{UdR Parma, INSTM, I-43124 Parma, Italy}
\affiliation{Gruppo Collegato di Parma, INFN-Sezione Milano-Bicocca, I-43124 Parma, Italy}

\begin{abstract}
Many problems intractable on classical devices could be solved by algorithms explicitly  based on quantum mechanical laws, i.e. exploiting quantum information processing. As a result, increasing efforts from different fields are nowadays directed to the actual realization of quantum devices. 
Here we provide an introduction to Quantum Information Processing, focusing on a promising setup for its implementation, represented by molecular spin clusters known as Molecular Nanomagnets. We introduce the basic tools to understand and design quantum algorithms, always referring to their actual realization on a molecular spin architecture. We then examine the most important sources of noise in this class of systems and then one of their most peculiar features, i.e. the possibility to exploit many (more than two) available states to encode information and to self-correct it from errors via proper design of quantum error correction codes. Finally, we present some examples of quantum algorithms proposed and implemented on a molecular spin qudit hardware.  \\
\\
This is an original manuscript of an article published by Taylor \& Francis in Contemporary Physics on 20th of August 2024, available online: \href{https://doi.org/10.1080/00107514.2024.2381952}{https://doi.org/10.1080/00107514.2024.2381952}.
\end{abstract}                              
\maketitle

\twocolumngrid 

\indent 

\section{Introduction}
Quantum computers promise to solve problems impossible to be addressed by classical digital devices.
Indeed, by encoding and processing information according to the laws of quantum mechanics, a single processor can 
perform many calculations in parallel, thus often overcoming the computational power of even the best classical super-computers. \\
However, the actual implementation of these powerful machines is not an easy task \cite{Feynman1982}. 
In principle, any quantum two-level system can be used to encode a quantum bit (qubit) of information. In practice, however, the operation of a quantum computer requires an impressive degree of control on the hardware to implement the sequence of gates needed to build up an algorithm. Hence, although the basic principles of quantum computation were already settled in the 90s, research on its physical implementation is still ongoing. So far, only a few architectures such as superconducting qubits and trapped ions \cite{Rev_SCqubits,RevTrappedIons,RevTacchino,RMPBlais} have reached  relatively high maturity through the efforts of large companies, while other less explored platforms could potentially offer important advantages \cite{Rev_NV,Rev_photonics}. 
Among these are spin systems, prototypical quantum objects which naturally provide two or more discrete energy levels which can form coherent superpositions and hence are suitable to encode the elementary units of quantum information.

Here we focus on Molecular Nanomagnets (MNMs), which represent a particularly promising class of spin qubits, and use them as a playground to illustrate basic principles of quantum computing. By targeting a specific hardware, we directly translate each computational step into its physical implementation, thus providing a sort of practical user-guide for those facing this field of research. Through the following sections, we address all the DiVincenzo criteria \cite{DiVincenzoC} required for the physical implementation of quantum computing on a given platform: (i) a scalable physical system with well-characterized qubit (Secs. \ref{sec:qubits},\ref{sec:2qubits}); (ii) the ability to initialise the state of the qubits into a simple fiducial state (\ref{sec:qubits},\ref{sec:readout}); (iii) long relevant decoherence times (\ref{sec:deco}); (iv) a "universal" set of quantum gates (\ref{sec:qubits},\ref{sec:2qubits}); (v) a qubit-specific measurement capability (\ref{sec:readout}).

MNMs are clusters of transition metal or rare-earth ions, coupled through organic ligands \cite{BOOK}. They can be synthesized in solutions, ordered in a crystal lattice or deposited onto metal/superconducting surfaces \cite{Atzori2016,Sessoli1993,Cornia2011}.
They are rather unique systems for essentially two reasons: (i) their spin Hamiltonian naturally provides many (more than two) low-energy levels which can be manipulated coherently by electromagnetic pulses. These additional states represent a crucial resource for many algorithm and in particular for quantum error correction. (ii) Their properties can be tailored at the synthetic level, thus achieving the required hierarchy of interactions in the spin Hamiltonian for specific implementations of quantum gates and protocols. \\
Many theoretical proposals have been put forward over the years for quantum information processing (QIP) with MNMs as elementary units. Recently, the focus has been on exploiting them as prototypical qudits (quantum information units with more than 2 levels), potentially embedding a qubit with quantum-error correction. Even if slower, recent progress has also been registered on the experimental side, with demonstration of two-qubit gates \cite{Ardavan2015}, the first proof-of-concept quantum simulator \cite{Chicco2023} and important steps towards addressing and reading-out the state of single molecules in superconducting resonators \cite{Gimeno2020}. 
Although harder to be scaled up, single-molecule transistors were realized a few years ago \cite{Thiele2014}, in which the state of a single molecule was accessed by measuring the current through a coupled conducting quantum dot \cite{Thiele2014} and a qudit version of the Grover's search algorithm was implemented \cite{Godfrin2017}.

The paper is organised as follows: we first illustrate the fundamental differences between classical and quantum computation
(Sec. \ref{sec:qubits}) in the encoding and processing of information. We then introduce multiple qubits and entangling operations in Sec. \ref{sec:2qubits}, thus identifying universal sets of one- and two-qubit gates, which form the building blocks of any algorithm. We then illustrate the harmful effects of decoherence, which turn quantum into classical information (Sec. \ref{sec:deco}) and how quantum-error correction can actually overcome it. Then we show how to extract the output of an algorithms by state readout (Sec. \ref{sec:readout}). In all sections, the mathematical formalism of quantum computing is introduced in parallel to the corresponding physical implementation in a molecular spin quantum hardware.
Having laid the foundations, we move to examples of more complex algorithms (Sec. \ref{sec:algorithms}) and we finally outline the specific features (Sec. \ref{sec:advantages}) which distinguish MNMs from other quantum computing platforms and could make them competitive with other technologies in the near future.

\section{Quantum vs classical bits: encoding and manipulation}
\label{sec:qubits}
\begin{figure*}[ht]
    \centering
	\includegraphics[width=0.9\textwidth]{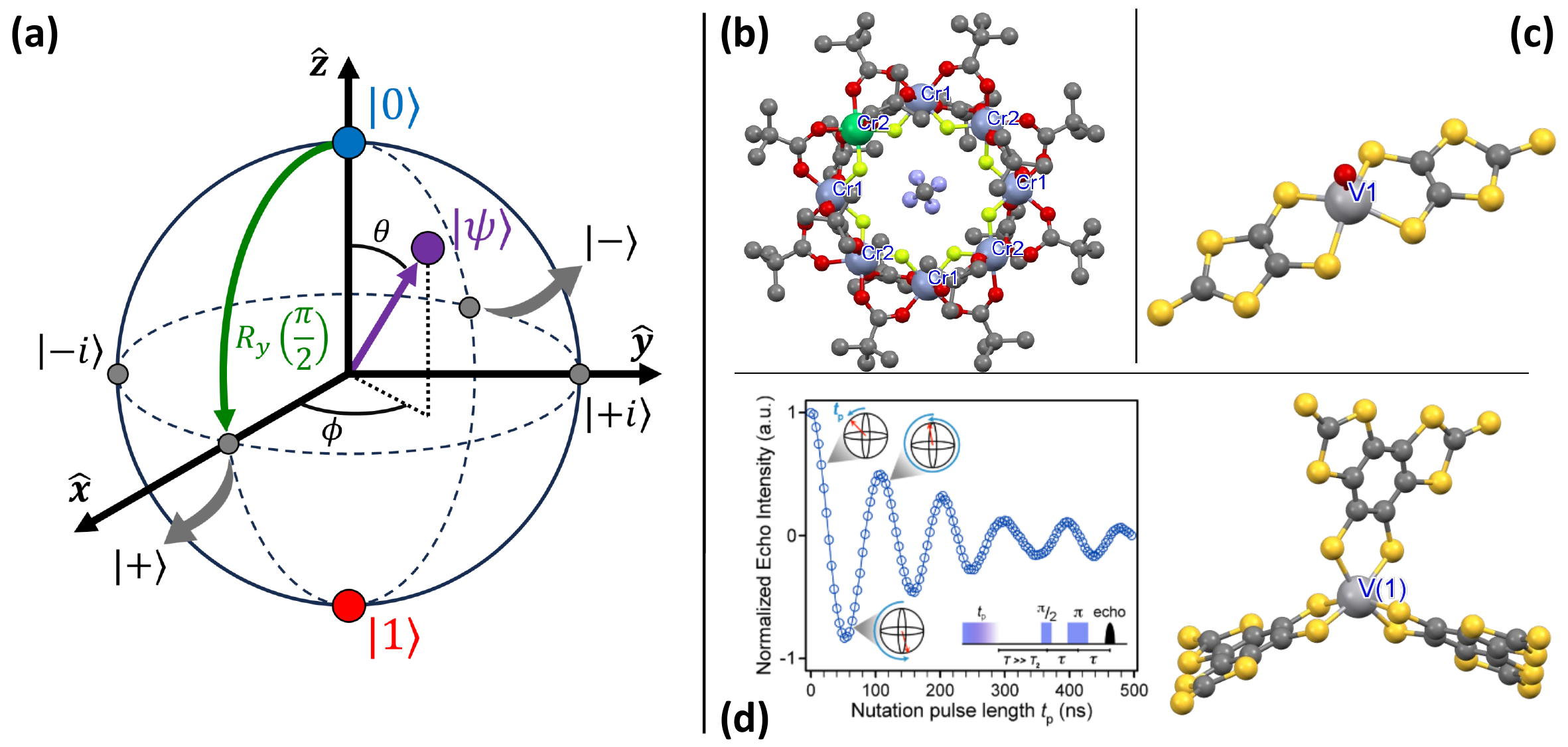}
	\caption{(a): Bloch sphere representation of a qubit state. We highlight in blue the north pole ($\ket{0}$), in red the south pole ($\ket{1}$). The purple dot represents a generic superposition $\ket{\psi}$, which lies on the surface of the sphere and hence can be described by the two angles $\theta,\phi$. The four gray points on the equator of the sphere represent the $\ket{\pm}$ and $\ket{\pm i}$ states. The green arrow shows the effect of a $\pi/2$ rotation about the $y$ axis, which moves the qubit state over the surface of the sphere from $\ket{0}$ to $\ket{+}$. (b-d) Structures of prototypical molecular qubits (with labeled transition metal ions): Cr$_7$Ni ring (b), a single-ion vanadium-based complex proposed in Ref. [\onlinecite{Atzori_JACS}] (c) and another vanadium-based compound along with the experimental detection of its Rabi oscillations, reprinted with permission from \cite{Zadrozny2014} (Copyright 2014 American Chemical Society).}
	\label{fig:bloch_sphere}
\end{figure*}

In perfect analogy with classical computation, QIP is usually based on a binary encoding of information.
This means that the elementary units of computation are two-level quantum systems called qubits, whose prototypical implementation is represented by a spin 1/2 particle. MNMs consisting of a single $s=1/2$ metal ion, such as VO \cite{Atzori2016,Atzori_JACS,Zadrozny2015} or Cu$^{2+}$ \cite{BaderNatComm14,Bader2016}, provide a clean example of such an implementation. When placed in an external static magnetic field $B$, the Zeeman interaction
\begin{equation}
	H_Z = g \mu_B B s_z
	\label{eq:Zeeman}
\end{equation}
splits the spin doublet into two distinct energy levels. Here $\mu_B$ is the Bohr magneton, $g$ is the spectroscopic factor and $s_z$ is the component of the spin operator parallel to ${\bf B}$ (assumed along $z$ for sake of simplicity).
The two logical states of the qubit, labelled as $\ket{0}$ and $\ket{1}$, can then be associated with the two spin projections along the static field, namely $\ket{\uparrow}\equiv\ket{0}$ and $\ket{\downarrow}\equiv \ket{1}$. 
Whereas a classical bit of information can only be in one of these two states (0 or 1), a quantum bit can exist in any superposition
\begin{equation}
	\ket{\psi} = \alpha \ket{0} + \beta \ket{1}
	\label{eq:superp}
\end{equation}
where $\alpha$ and $\beta$ are complex coefficients normalised to 1, i.e. $|\alpha|^2+|\beta|^2=1$.
A useful representation of the generic state of a qubit is given by the Bloch sphere (see Fig. \ref{fig:bloch_sphere}-(a)). Indeed, if we set $\alpha= \cos \frac{\theta}{2}$ and $\beta=\sin \frac{\theta}{2}  e^{-i\varphi}$, a generic state of the qubit is represented (up to an overall phase), by a point onto the surface of a sphere of unitary radius (the so-called Bloch sphere), where $\theta$ and $\varphi$ are the polar and azimuthal angles.
The North and the South poles (along the $z$ axis), associated to the polar angles $\theta=0$ and $\theta=\pi$, identify the "classical" states $\ket{0}$ and $\ket{1}$, i.e. the eigenstates of $\sigma_z$ Pauli matrix.
By varying $\theta$ one can generate superpositions with different weights of $\alpha$ and $\beta$. In particular, eigenstates of the other Pauli matrices lie on the equator of the Bloch sphere, corresponding to $\theta=\pi/2$. For $\varphi=0,\pi$ we get the eigenstates of $\sigma_x$, aligned along $x$ axis 
\begin{equation}
	\ket{\pm}=\frac{\ket{0}\pm\ket{1}}{\sqrt{2}}
\end{equation}
while for $\varphi=\pm\pi/2$ the corresponding eigenstates of $\sigma_y$, along $y$ axis:
\begin{equation}
	\ket{\pm i}=\frac{\ket{0}\pm i\ket{1}}{\sqrt{2}}.
\end{equation}

Remarkable realizations of qubits based on MNMs are single- or multi-spin clusters characterized by an (effective) spin 1/2 ground state. The forefather is probably represented by Cr$_7$Ni ring \cite{Larsen2003,Troiani2005,Troiani2005b} [see Fig. \ref{fig:bloch_sphere}-(b)], consisting of seven spin 3/2 Cr$^{3+}$ and a spin 1 Ni$^{2+}$ ion with rather strong nearest-neighbors anti-ferromagnetic coupling. This gives rise to an effective spin 1/2 ground doublet which can be split by an external field and used to encode a qubit as described above. A prominent feature of this molecule is the high degree of chemical control, which allowed Chemists to (i) improve its coherence times by making the structure more rigid and hence less subject to low-energy vibrations \cite{Wedge2012}; (ii) anchor it onto surfaces with addition of proper ligands \cite{Corradini2012}; (iii) combine with other molecular qubits into complex supra-molecular multi-qubit structures with tailored interactions \cite{Timco2009,Lockyer2022}.   \\
As discussed in more detail in Sec. \ref{sec:deco}, increasing the coherence time $T_2$ is one of the most important goals in the synthesis of suitable molecular qubits. Roughly speaking, the coherence time sets the longest timescale up to which quantum superpositions exist and hence quantum information can be considered reliable. The longer $T_2$, the larger the number of operations which can be implemented on the qubit, before it collapses onto a classical state. Along the years the 1-10 $\mu$s values of $T_2$ of Cr$_7$Ni have been significantly enhanced by simplifying the molecular structures \cite{Moro2014}. In particular, single-ion molecules based on Cu$^{2+}$ or V [see Fig. \ref{fig:bloch_sphere}-(c,d)] have reached $T_2$ up to 70 $\mu$s and even close to ms \cite{BaderNatComm14,Zadrozny2015,Freedman_JACS}, preserving in some cases significant values also at room temperature \cite{Atzori2016,Bader2016,Atzori_JACS}.

\subsection{Single-qubit operations} 
\label{sec:rotations}
From the above introduction it is already evident that linear algebra is at the very core of QIP: 
any state of the computational basis is a vector in a linear vector space called Hilbert space. 
Quantum operations (gates) on a single qubit have a similarly intuitive geometric interpretation: since the qubit space is a linear vector space, any gate must keep the state within such a linear space. Hence, 
quantum gates are operators which can be represented as unitary matrices.
This implies that {\it ideal} single-qubit gates rotate the state of the qubit over the surface of the Bloch sphere. Let us consider for instance rotations about the three orthogonal $x,y$ and $z$ axes. The corresponding operator can be written as:
\begin{equation}
	R_\alpha(\theta) = e^{-i \sigma_\alpha \theta/2} = \cos \frac{\theta}{2} I -i \sin \frac{\theta}{2} \sigma_\alpha,
	\label{eq:rotation}
\end{equation}
where $\sigma_\alpha = 2 s_\alpha$ ($\alpha=x,y,z$) are Pauli matrices and we have exploited known identities ($\sigma^{2n}_\alpha = I$, $\sigma^{2n+1}_\alpha = \sigma_\alpha$) to derive the last expression \cite{Nielsen}.
Rotations about $x$ or $y$ [such as the green arrow in Fig. \ref{fig:bloch_sphere}-(a)] correspond to varying the polar angle $\theta$ which identifies the state onto the Bloch sphere, while rotations about $z$ introduce a relative phase between $\ket{0}$ and $\ket{1}$ and hence rotate the wavefunction in the plane orthogonal to $z$. 
Note that, in spite of this simple geometric interpretation, a $2\pi$ rotation does not correspond to the identity operator, but to $-I$ and hence it adds an overall $\pi$ phase to the single-qubit wavefunction. This 
is a purely quantum-mechanical effect arising from the $4\pi$ periodicity of spin 1/2 systems, which can be exploited to construct non trivial gates \cite{modules,SciRepNi,Chizzini2022} (see below).

We now consider how to implement these rotations on a spin 1/2 molecule. 
In a magnetic field applied along the $z$ quantization axis, the computational basis states $\ket{0} \equiv \ket{\uparrow}$ and $\ket{1} \equiv \ket{\downarrow}$ are eigenstates of Hamiltonian \eqref{eq:Zeeman}, split by an energy gap $\Delta E = g \mu_B B$.
If we prepare the system in $\ket{\downarrow}$, we can exploit a resonant transverse microwave pulse at angular frequency $\omega = \Delta E/\hbar$ to induce a transition from $\ket{\downarrow}$ to $\ket{\uparrow}$. 
The time-dependent Hamiltonian of this driving field can be expressed as
\begin{eqnarray}  \nonumber
	H_1(t) &=& g_\perp \mu_B B_1
	\Theta(|t-t_0|-\tau/2) \\
	&&\left[ s_x \cos (\omega t + \phi) + s_y \sin (\omega t + \phi) \right]  ,
	\label{eq:pulse}
\end{eqnarray}
where for simplicity we have assumed a circularly polarized rectangular-shaped pulse of center $t_0$, duration $\tau$, frequency $\omega/2\pi$, with ${\bf B}_1$ oriented perpendicular to $z$ ($g_\perp$ is the transversal component of the spectroscopic tensor ${\bf g}$). 
This pulse implements a $R_{\bf n}(\theta)$ rotation, in which ${\bf n}$ is a versor in the $xy$ plane.
The specific orientation of ${\bf n}$ can be controlled by the phase $\phi$ in Eq. \eqref{eq:pulse}, while the duration $\tau$ of the pulse sets the rotation angle $\theta$. This can be easily understood by re-writing the total Hamiltonian $H_Z+H_1(t)$ in a frame rotating about $z$ at frequency $\omega/2\pi$, i.e. applying the transformation $R=\exp{i \omega t s_z}$.
Then (for $t_0-\tau/2<t<t_0+\tau/2$) we get
\begin{eqnarray} 
	H_{RF} &=& R^\dagger [H_Z+H_1(t)] R - i \hbar \dot{R}^\dagger R \\ \nonumber
	&=& (\Delta E -\hbar \omega) s_z + g_\perp \mu_B B_1 \left( s_x \cos \phi + s_y \sin \phi \right) \\ \nonumber
	&=& \delta s_z + \gamma \left( s_x \cos \phi + s_y \sin \phi \right) .
\end{eqnarray}
where we have defined the frequency de-tuning parameter $\delta=g \mu_B B -\hbar \omega$ and $\gamma=g_\perp \mu_B B_1$. Note that we have effectively removed explicit time dependence from the Hamiltonian by writing it in the rotating frame, where time evolution is ruled by the operator $U(t)= e^{-i H_{RF} t/\hbar}$.
It is now clear that if we use the resonant frequency $\omega = \Delta E/\hbar$ ($\delta=0$) and we let the system evolve for a time $\tau=\hbar \theta/\gamma$, we implement a generic planar rotation about an axis in the $xy$ plane determined by $\phi$. For instance,  $U(\tau)=R_x(\theta)$ for $\phi=0$ [green arrow in Fig. \ref{fig:bloch_sphere}-(a)] or $U(\tau)=R_y(\theta)$ for $\phi=\pi/2$. 
Moving back to the laboratory frame only changes the relative phase between $\ket{0}$ and $\ket{1}$, which is usually accounted for at the software level. \\
If we instead use an out of resonance frequency ($\delta\neq0$), only a fraction $\sim \gamma/\delta$ of the wavefunction evolves between basis states $\ket{0}$ and $\ket{1}$.
In particular, in the so-called semi-resonant regime, a pulse lasting $2\pi/\sqrt{\delta^2+4\gamma^2}$ can be exploited to induce a partial transfer of population from $\ket{0}$ to $\ket{1}$ and then back to $\ket{0}$. By properly choosing $\delta$, this results in a relative phase $\varphi = \pi \left[ 1 -  \delta/\sqrt{\delta^2 + 4\gamma^2}\right]$ between the two states after the pulse, which corresponds to implementing a $R_z(\varphi)$ rotation\cite{Mariantoni2011,SciRep15,prb2016,Chiesa2023prappl}. 

In general, rotations of an arbitrary angle about  two non-parallel axes of the Bloch sphere are sufficient to build a universal single-qubit gate \cite{Nielsen}. In fact, a generic single-quit gate represented by the unitary operator $U \in SU(2)$ can be decomposed in a sequence of $y$ and $z$ rotations as $U=\exp{i\alpha}R_z(\beta)R_y(\gamma)R_z(\delta)$  \cite{Nielsen,Sakurai}.\\
Unitary operators, i.e. operators satisfying $U^\dagger U = U U^\dagger = I$, preserve scalar products between different qubit states. 
In the Bloch sphere representation, this {\it unitarity} implies that we keep on moving onto its surface. 
A further important consequence of the unitary quantum dynamics is its {\it reversibility}. This means that after one or a sequence of quantum gates 
represented by $U$ we can restore the initial state in a deterministic way by simply applying the gate $U^{-1} = U^\dagger$. This property also holds for multi-qubit gates, which are always represented by unitary operators, and is in contrast with classical computation, where two-bit operations can map a 2-bit input into a single-bit output, thus hindering the univocal inversion of the operation (see Sec. \ref{sec:2qubits} below). \\
As we will see in Sec. \ref{sec:deco}, the interaction between qubits and environment breaks the unitarity of the dynamics, thus making it irreversible and causing errors in the implementation of quantum algorithms. At the single-qubit level, this harmful irreversible dynamics can be seen as an effective shrinking of the Bloch sphere \cite{Nielsen}.

Before concluding this section, we introduce an important single qubit gate, the Hadamard gate.
In the computational $\{\ket{0},\ket{1}\}$ basis, it is represented by the matrix
\begin{equation}
	H_d =   
	\begin{pmatrix}
		1 &  1 \\
		1 & -1
	\end{pmatrix}.
	\label{eq:Hadamard}
\end{equation}
This transformation maps the eigenstates of $\sigma_z$ into the eigenstates of $\sigma_x$, i.e. 
$\ket{0}\rightarrow(\ket{0}+\ket{1})/\sqrt{2}\equiv \ket{+}$ and $\ket{1}\rightarrow(\ket{0}-\ket{1})/\sqrt{2} \equiv\ket{-}$. \\
\begin{figure}
	\includegraphics[width=0.48\textwidth]{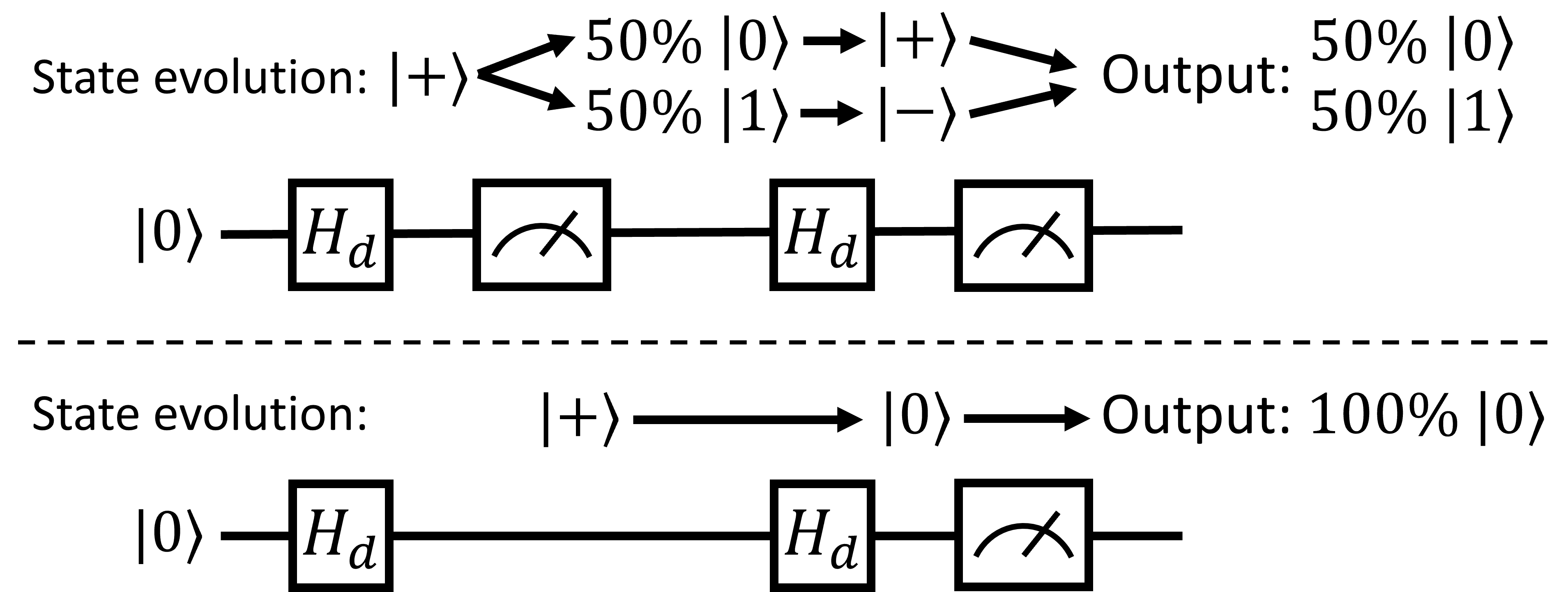}
	\caption{A one-qubit quantum circuit without (top) or with (bottom) quantum interference. Over each circuit we highlight the evolution of the state of the qubit and the probabilities associated with the possible results of each measurement. }
	\label{fig:interference}
\end{figure}
We can exploit the Hadamard gate to introduce an important concept in QIP which makes it different from classical one, i.e. {\it interference} \cite{KayeBOOK}. In order to delve into this topic, let us introduce the two single-qubit quantum circuits shown in Fig. \ref{fig:interference}. Quantum circuits are used to aid the visual representation of the sequence of operations on one or more qubits: each line of the circuit represents the state of the qubit over time (from left to right). At their left they are labeled with the initial state of that qubit, whereas squares placed on the lines represent single-qubit quantum gates and measurements, each one with a specific symbol. As an example, in Fig. \ref{fig:interference} the qubit is initialised in $\ket{0}$ in both cases and is subject to a sequence of Hadamard gates and measurements in the $\{\ket{0},\ket{1} \}$ basis.
We send to Sec. \ref{sec:readout} for more details on qubit state readout. Here we simply consider its effect as a projection onto either $\ket{0}$ or $\ket{1}$ with probability given by $|\alpha|^2$ or $|\beta|^2$ in Eq. \eqref{eq:superp}. 
In both circuits, the initial qubit state $\ket{0}$ is brought by $H_d$ into $\ket{+}=(\ket{0}+\ket{1})/\sqrt{2}$. Then, if we measure the state of the qubit (top circuit), the superposition state is collapsed either to $\ket{0}$ or $\ket{1}$ with equal $1/2$ probabilities. Depending on the measurement outcome, the subsequent $H_d$ gate produces either $\ket{+}$ or $\ket{-}$, both of which in the final measurement yield $\ket{0}$ or $\ket{1}$ with 50\% probability. This corresponds to a classical probabilistic circuit, in which we have checked the state of the qubit in between the two $H_d$ gates, thus in fact destroying the superposition. Instead, if we remove the intermediate measurement (bottom circuit), the $\ket{+}$ state is brought back to $\ket{0}$ by $H_d$ and hence this is the only possible output of the final measurement. In other words, the quantum circuit has produced a sort of {\it constructive interference} for $\ket{0}$ and {\it destructive} for $\ket{1}$. \\
These interference paths are a very important tool to design quantum algorithms and follow, again, from the linearity of quantum gates. In contrast, measurements introduce non-linear effects which cause the collapse of the quantum state into one of its eigenstates, thus in fact turning information from quantum into classical. Hence, detection of the state of one or more qubits is not as innocent as in classical computing, because it can dramatically affect the system state.

\section{Multiple qubits and entanglement}
\label{sec:2qubits}
The building blocks introduced above must then be combined into a {\it multi-qubit register} to form a quantum hardware.
The simplest implementation of this register is given by a collection of $N$ non-interacting qubits. A basis for the corresponding
 $2^N$ dimensional Hilbert space is given by {\it factorized} (product) states of the form $\ket{\psi_1}\otimes \dots \otimes \ket{\psi_N} \equiv \ket{\psi_1 \dots \psi_N}$, where $\ket{\psi_i}$ is the state of each single qubit. 
This means that any multi-qubit quantum state in this Hilbert space can be expressed as
 \begin{equation}
 	\ket{\psi} = \sum_k \alpha_k \ket{\psi_1^k \dots \otimes \psi_N^k}.
 	\label{eq:productExp}
 \end{equation}
This is a superposition of product states but for a generic choice of $\alpha_k$ it is no longer factorized. The impossibility of factorizing a multi-qubit state leads to correlations between measurement results on different qubits. This property is called {\it entanglement} and is, together with {\it superposition} and {\it interference}, one of the strictly-quantum ingredients of quantum computation. \\
Let us explore this aspect by considering a 2-qubit system in the state $\ket{\psi_0}=\ket{11}$, realized for instance by a pair of
spins 1/2 ions in a magnetic field, initialized in their ground state $\ket{\downarrow \downarrow}$.
As a first step, we apply a Hadamard gate only to the first qubit of the register, obtaining $\ket{\psi_1} = \ket{-} \otimes \ket{1}$. This state is still factorized, because we can exactly say that the first qubit is in an eigenstate of $\sigma_x$, while the second is in an eigenstate of $\sigma_z$.
We now apply to $\ket{\psi_1}$ a two-qubit gate represented, in the computational $\{ \ket{00},\ket{01},\ket{10},\ket{11} \}$ basis, by the matrix
\begin{equation}
	U_{cX} = 
	\begin{pmatrix}
		1 & 0 & 0 & 0  \\
		0 & 1 & 0 & 0 \\
		0 & 0 & 0 & 1 \\
		0 & 0 & 1 & 0
	\end{pmatrix}
\label{eq:cnot}
\end{equation}
which is known as controlled-NOT or cX gate. The implementation of this gate leaves components $\ket{00}$ and $\ket{01}$ of a generic two-qubit state unaffected, while it exchanges $\ket{10}$ and $\ket{11}$. It is a {\it controlled} operation, meaning that it applies a gate to the second ({\it target}) qubit iff the first qubit (called {\it control}) is in $\ket{1}$. In the specific case of the cX, the gate applied to the target is $\sigma_x$, which flips the state of the qubit ($\ket{0} \leftrightarrow\ket{1}$) analogously to the classical NOT. In spite of this apparent simplicity, when applied to $\ket{\psi_1}$ the cX 
generates 
\begin{equation}
    \ket{\Psi_-} = U_{cX}\frac{\ket{0}-\ket{1}}{\sqrt{2}} \otimes \ket{1} = \frac{\ket{01}-\ket{10}}{\sqrt{2}},
\end{equation}
which is no longer factorized but {\it entangled}, because we cannot specify the state of the first or of the second qubit alone. 
To clarify this point, let us consider the measurement of a local observable $M$, acting only on one of the two qubits (e.g. the first one).
This corresponds to compute the expectation value $\bra{\Psi_-} M \otimes I \ket{\Psi_-}$.
By simple algebra we find:
\begin{equation}
	\bra{\Psi_-} M \otimes I \ket{\Psi_-} = \frac{1}{2} \left( \bra{0} M \ket{0} + \bra{1} M \ket{1} \right).
	\label{eq:psimmix}
\end{equation}
Notably, there is no single-qubit state vector $\ket{\psi} = \alpha \ket{0} + \beta \ket{1}$ which can produce this result in the measurement of observable $M$. For instance, $\bra{+} M \ket{+} = (\bra{0} M \ket{0} + \bra{1} M \ket{1} + \bra{0} M \ket{1} + \bra{1} M \ket{0})/2$, which is different from Eq. \eqref{eq:psimmix}. Hence, a measurement of a local observable on an entangled pair collapses the whole two-qubit state into an incoherent mixture (see details in Sec. \ref{sec:deco}), thus affecting both qubits simultaneously. \\
As a matter of fact, $\ket{\Psi_-}$ is one of the four maximally entangled two-qubit states known as Bell states (there are metrics to quantify the amount of entanglement such as the concurrence or the negativity, see e.g. \cite{Nielsen}).
The full list of Bell states is the following:
\begin{eqnarray}
\ket{\Psi_+} &=& \frac{\ket{01}+\ket{10}}{\sqrt{2}} \equiv U_{cX} \ket{+}\otimes \ket{1} \\
\ket{\Psi_-} &=& \frac{\ket{01}-\ket{10}}{\sqrt{2}} \equiv U_{cX} \ket{-}\otimes \ket{1} \\
\ket{\Phi_+} &=& \frac{\ket{00}+\ket{11}}{\sqrt{2}} \equiv U_{cX} \ket{+}\otimes \ket{0} \\
\ket{\Phi_+} &=& \frac{\ket{00}-\ket{11}}{\sqrt{2}} \equiv U_{cX} \ket{-}\otimes \ket{0} ,
\end{eqnarray}
which can all be obtained from a cX gate applied to a factorized state, as shown by the last identities.
Thanks to its capability of transforming factorized into maximally-entangled Bell states, the cX is very important in quantum computation.
In fact, it can be shown that arbitrary single-qubit rotations on all the qubits of the register, combined with a single type of entangling two-qubit gates (such as the cX) between all the qubit pairs in the register form a universal set, i.e. any multi-qubit operation can be decomposed using these elementary operations \cite{Nielsen}.\\
The cX has some similarities with the classical XOR gate. The latter is an irreversible gate, which starting from a pair of bits in the generic state $ab$ produces a single output bit given by $a \oplus b$ (here $\oplus$ is the sum modulo 2). Clearly, it is impossible to reconstruct the state of the input from the single qubit we get as an output. However, the {\it reversible} cX gate can exactly be written as a XOR on the target qubit with the control unaltered, i.e. $\ket{ab} \leftrightarrow \ket{a,a\oplus b}$. 
This trick of adding a control register is commonly used in QIP (where operations are always unitary and reversible) to compute {\it irreversible} functions on the target \cite{Nielsen,Barnett}. \\

\subsection{Two-qubit gates in molecular spins}\label{switch}
\begin{figure}[t]
	\includegraphics[width=0.48\textwidth]{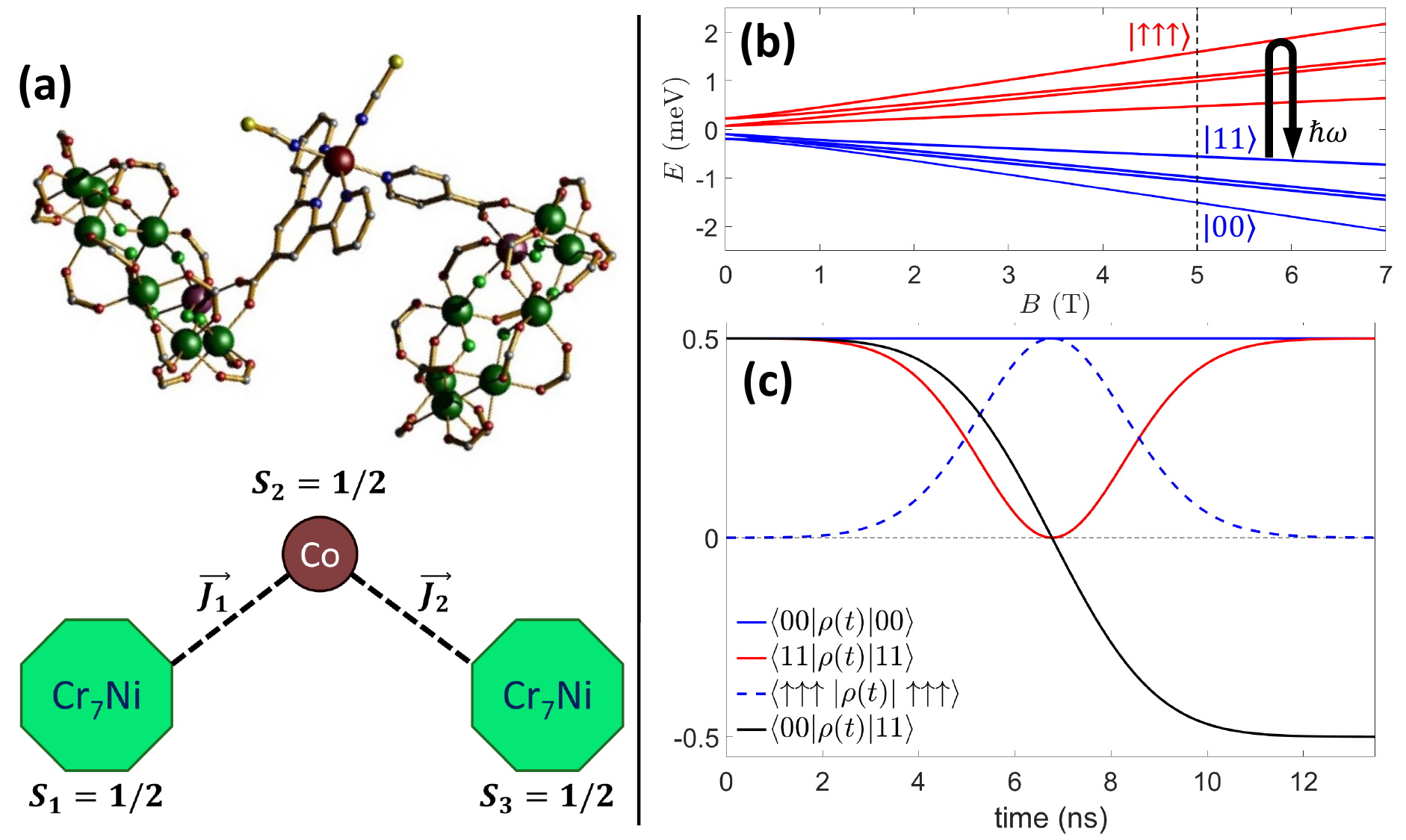}
	\caption{(a) Molecular structure of one of the Cr$_7$Ni-Co-Cr$_7$Ni $S=1/2$ trimers proposed in [\onlinecite{modules}] for the implementation of two-qubit gates via a switchable interaction. Bottom: scheme of the molecules, with nearest-neighbor exchange couplings ${\bf J}_{1,2}$ reported. Molecular structure adapted with permission from \cite{modules}. (b) Energy levels of the molecule represented in panel (a) as a function of static magnetic field $B$, with parameters derived from electron-paramagnetic resonance measurements \cite{modules}: $g_{1,3}=(1.74,1.78,1.78)$, $g_2=(2,4.25,6.5)$, $\mathbf{J}_1=(-0.14,0.34,0.17)$ cm$^{-1}$ and $\mathbf{J}_2=(-0.07,0.17,0.34)$ cm$^{-1}$. The black arrow represents the $2\pi$ pulse resonant with the $\ket{\uparrow\downarrow\uparrow}\leftrightarrow\ket{\uparrow\uparrow\uparrow}$ transition required to implement a cZ gate as described in section \ref{switch}. (c) Simulated time-evolution of the system state during the implementation of the cZ gate. Solid blue and red lines represent the population over time of the computation basis states $\ket{00}$ and $\ket{11}$. The dashed blue line represents instead the population over time of the auxiliary state $\ket{\uparrow\uparrow\uparrow}$. The solid black line represents the coherence between $\ket{00}$ and $\ket{11}$. }
	\label{fig:cz_switch}
\end{figure}

Depending on the specific implementation, entanglement can be introduced also by employing two-qubit gates other than the cX.
Indeed, the paradigmatic cX gate represents only an example of controlled (c$U$) operations, in which the state of the target is transformed according to a generic single-qubit rotation $U$ for a specific state of the control.
In the case of MNMs, different schemes have been proposed. One of the most promising relies on the c$\varphi$ gate, i.e. a gate diagonal in the computational basis, represented by the matrix
\begin{equation}
	U_{c\varphi} = 
	\begin{pmatrix}
		1 & 0 & 0 & 0 \\
		0 & 1 & 0 & 0 \\
		0 & 0 & 1 & 0 \\
		0 & 0 & 0 & e^{-i\varphi} 
	\end{pmatrix}
\label{eq:cphi}
\end{equation}
which adds a phase $e^{-i\varphi}$ only to $\ket{11}$ component.
This gate can be implemented by considering a trimer of three spins 1/2 such as the one sketched in Fig. \ref{fig:cz_switch}-(a), and described by the spin Hamiltonian:
\begin{eqnarray} \nonumber
	H_3 &=& \mu_B B \sum_{k=1}^3 g_k s_{k}^z + \sum_{k=1}^2 J_k^z s_k^z s_{k+1}^z \\
	&+& \sum_{k=1}^2 J_k^\perp (s_k^x s_{k+1}^x + s_k^y s_{k+1}^y).
	\label{eq:SHtrimer}
\end{eqnarray}
Besides the Zeeman interaction with an external magnetic field parallel to $z$, here we have introduced an interaction between neighboring sites in the three-spin chain, with both longitudinal ($J^z_k$) and perpendicular ($J^\perp_k$) contributions.
For $J^\perp_k=0$, the eigenstates of Hamiltonian \eqref{eq:SHtrimer} are factorized and correspond to the eigenstates of $s^z_k$. Hence, they can be used to define a proper 2-qubit register, in which the possibility of implementing a two-qubit entangling gate must be provided. To achieve this,  we encode the logical states of the qubit pair into spins 1 and 3, while spin 2 acts as a dynamical switch of the effective qubit-qubit interaction between qubits 1 and 3. Hence, the computational subspace is defined with spin 2 frozen into its $\ket{\downarrow}$ state, i.e. $\ket{00} \equiv\ket{\uparrow \downarrow \uparrow}, \ket{01} \equiv\ket{\uparrow \downarrow \downarrow}, \ket{10} \equiv\ket{\downarrow \downarrow \uparrow}, \ket{11} \equiv\ket{\downarrow \downarrow \downarrow}$. 
The possibility of turning on and off the inter-qubit coupling is crucial for the proper operation of the quantum hardware. Indeed, in a generic algorithm one needs to alternate single-qubit gates, idle stages and two-qubit gates, with the inter-qubit coupling only active during the implementation of the latter. Otherwise, single-qubit operations will be imperfect and in general the multi-qubit state will be subject to unwanted evolutions which require complex and often unfeasible compensations.
In the molecular trimer shown in Fig. \ref{fig:cz_switch}-(a), the coupling between the three spins is permanent (it is set at the synthetic level), but we can nevertheless design a pulse sequence to effectively induce a switchable interaction between spins 1 and 3. \\
In particular, as long as the switch is frozen in $\ket{\downarrow}$, it only re-normalizes the effective magnetic field felt by the qubits, i.e. it acts as a one-body term which does not produce any effective coupling between the logical qubits.
However, excitation energies of the switch depend on the state of both qubits via the $J_k^z s_k^z s_{k+1}^z$ term in Hamiltonian \eqref{eq:SHtrimer}. In other words, 
implementing the transition $\ket{m_1 \downarrow m_3} \rightarrow \ket{m_1 \uparrow m_3}$ by a resonant pulse requires to choose the driving frequency according to
\begin{equation}
	\hbar \omega_{m_1m_3} = g_2 \mu_B B + (J^z_1 m_1 + J^z_2 m_3) .
\end{equation}
Since this frequency depends on the state $m_{1,3}=\ket{ \uparrow}/\ket{\downarrow}$ of both logical qubits, it effectively implements a conditioned evolution on specific two-qubit states.
In particular, a resonant $2\pi$ pulse on the switch performed only for $m_1=m_3=\ket{\uparrow}$ realizes a cZ gate on the qubits. Indeed, the $\ket{11}\equiv\ket{\uparrow\uparrow}$ component of the two-qubit state acquires a $\pi$ phase, while all the others are unaffected by the pulse, since they are off-resonant thanks to the $J^z_k$ couplings. 
The approach can be generalized to implement a c$\varphi$ gate, by addressing the switch via a semi-resonant pulse, detuned from resonance by a small amount $\delta$. As already outlined in Sec. \ref{sec:qubits}, a $2\pi$ semi-resonant pulse (which begins and ends with the switch in $\ket{\downarrow}$) can implement an arbitrary phase $\varphi$ depending on $\delta$, as desired \cite{Mariantoni2011,SciRep15}. 

So far we have neglected transverse terms in the spin Hamiltonian \eqref{eq:SHtrimer}, which however are in general present in real molecules. Small values of $J^\perp_k$ are still compatible with the scheme presented above. 
To achieve this, we need to fulfill the condition $J^\perp_k \ll |g_{1,3}-g_2| \mu_B B$. If this holds, the eigenstates will be only slightly different from the factorized computational basis. The transverse terms in the Hamiltonian introduce an effective coupling between the two qubits, but the corresponding (unwanted) evolution will occur on a timescale long compared to the pulses implementing the gates (each lasting typically tens of ns). Note that choosing non-equivalent qubits $g_1\neq g_3$ significantly helps to suppress such a dynamics. \\
Moreover, this evolution could be completely cancelled if the switch is realised by an anti-ferromagnetically coupled spin 1/2 dimer, characterised by a total spin $S=0$ ground state \cite{Santini2011}. Indeed, if the isotropic exchange is the leading interaction in the dimer, the total electronic spin $S=0,1$ is a good quantum number and the 4 eigenstates of the dimer are split into a singlet and a higher energy triplet. 
As far as the switch remains in the $S=0$ state the interaction is exactly turned off, while excitation to one of the $S=1$ states dynamically activate the 2-qubit c$\varphi$ gate as illustrated above. Although slower, microwave transitions between different total spin states of the switch are allowed in presence of different $g$ values on the two ions. \\
The simulation of a c$\varphi$ gate implemented on a real multi-qubit molecule is reported in Fig. \ref{fig:cz_switch}, following the procedure outlined above. The two logical qubits are Cr$_7$Ni rings, linked via an interposed Co$^{2+}$ complex which acts as a switch of their effective coupling [panel (a)]. The almost perpendicular arrangement of the two rings makes them inequivalent (for a generic orientation of ${\bf B}$) thanks to the anisotropy of their ${\bf g}$ tensors. This on the one hand suppresses the unwanted second-order evolution ruled by transverse couplings $J_k^\perp$. On the other hand, it allows one to individually address each qubit, because their respective transition frequencies are different. As a result, the symmetric c$\varphi$ gate can be transformed into a cX gate. This can be done by implementing a Hadamard gate {\it only on the target qubit} before and after the c$\varphi$ \cite{modules}, as follows from the equality $H_d \sigma_z H_d = \sigma_x$.

Alternative proposals for two qubit gates in molecular spins also exist. For instance, the switch can be provided by a redox-active unit, which can be brought from a diamagnetic to a spin 1/2 state by injecting an electron. Clearly, placing a diamagnetic ion at site 2 in the trimer shown in Fig. \ref{fig:cz_switch} corresponds to removing its spin from the Hamiltonian, thus cancelling the interaction between the logical qubits. If then a spin polarised electron is injected on site 2 (e.g. by an STM tip), the Hamiltonian becomes Eq. \eqref{eq:SHtrimer}, where however we consider $g_1=g_3$ and $J^z_k=J^\perp_k = J$. The residual interaction described above can now be exploited as a resource. In the limit $|g_1-g_2|\mu_B B \gg J$, we can consider the Zeeman term as the leading contribution to Hamiltonian \eqref{eq:SHtrimer} ($H_0$) and the exchange coupling as a perturbation $H_1$. Then, the free system dynamics is reduced to the computational subspace, in which the switch is $\ket{\downarrow}$ and the effect of $H_1$ is computed perturbatively by accounting for Virtual excitations of the switch to the $\ket{\uparrow}$ state. In other words, the effective 4x4 Hamiltonian in the computational subspace is given by
\begin{eqnarray} \nonumber
    \bra{i j} H_{eff} && \ket{i^\prime j^\prime} = \bra{i \downarrow j} H_{0} \ket{i^\prime \downarrow j^\prime} +   \\ \nonumber &&
    \sum_{i^{\prime\prime}j^{\prime\prime}} 
    \bra{i \downarrow  j}  H_1 \ket{i^{\prime\prime} \uparrow j^{\prime\prime}} \bra{i^{\prime\prime} \uparrow j^{\prime\prime}} H_1 \ket{i^\prime \downarrow j^\prime}  \\ &&
    \frac{1}{2} \left( \frac{1}{E_{ij}-E_{i^{\prime\prime}j^{\prime\prime}}} + \frac{1}{E_{i^\prime j^\prime}-E_{i^{\prime\prime}j^{\prime\prime}}} \right)
\end{eqnarray}
with $i,j,i^\prime, j^\prime,i^{\prime\prime},j^{\prime\prime} = \uparrow, \downarrow$. In practice, $H_0$ provides a first order splitting, while $H_1$ 
activates an effective interaction between the logical qubits \cite{modules,Chem2016} of the form
\begin{equation}
	H_{eff} = \Gamma (s_{x1} s_{x2}  + s_{y1} s_{y2} ) ,
\label{eq:Heff}
\end{equation}
with $\Gamma = J^2/2|g_1-g_2|\mu_B B$. 
If this interaction is left active for a time interval $\tau$, the corresponding unitary dynamics is
\begin{equation}
	U_{XY} (\tau) = e^{-i H_{eff}\tau} = 
	\begin{pmatrix}
		1 & 0 & 0 & 0 \\
		0 & \cos \Gamma \tau/2 & i \sin \Gamma \tau/2 & 0 \\
		0 & i\sin \Gamma \tau/2 & \cos \Gamma \tau/2 & 0 \\
		0 & 0 & 0 & 1		
	\end{pmatrix}
\label{eq:Uxy}
\end{equation}
By choosing $\tau = \pi/2\Gamma$, the resulting gate is the $\sqrt{\rm iSWAP}$, which
maps $\ket{01}$ to $(\ket{01}+i\ket{10})\sqrt{2}$ and $\ket{10}$ to $(\ket{10}+i\ket{01})\sqrt{2}$ and 
has entangling properties similar to the cX. Moreover, thanks to the flexibility in the choice of $\tau$, $U_{XY}(\tau)$ can provide efficient decomposition of several algorithms, especially for quantum simulations (see below).

\subsection{Scalability}
\begin{figure*}[ht]
    \centering
	\includegraphics[width=0.9\textwidth]{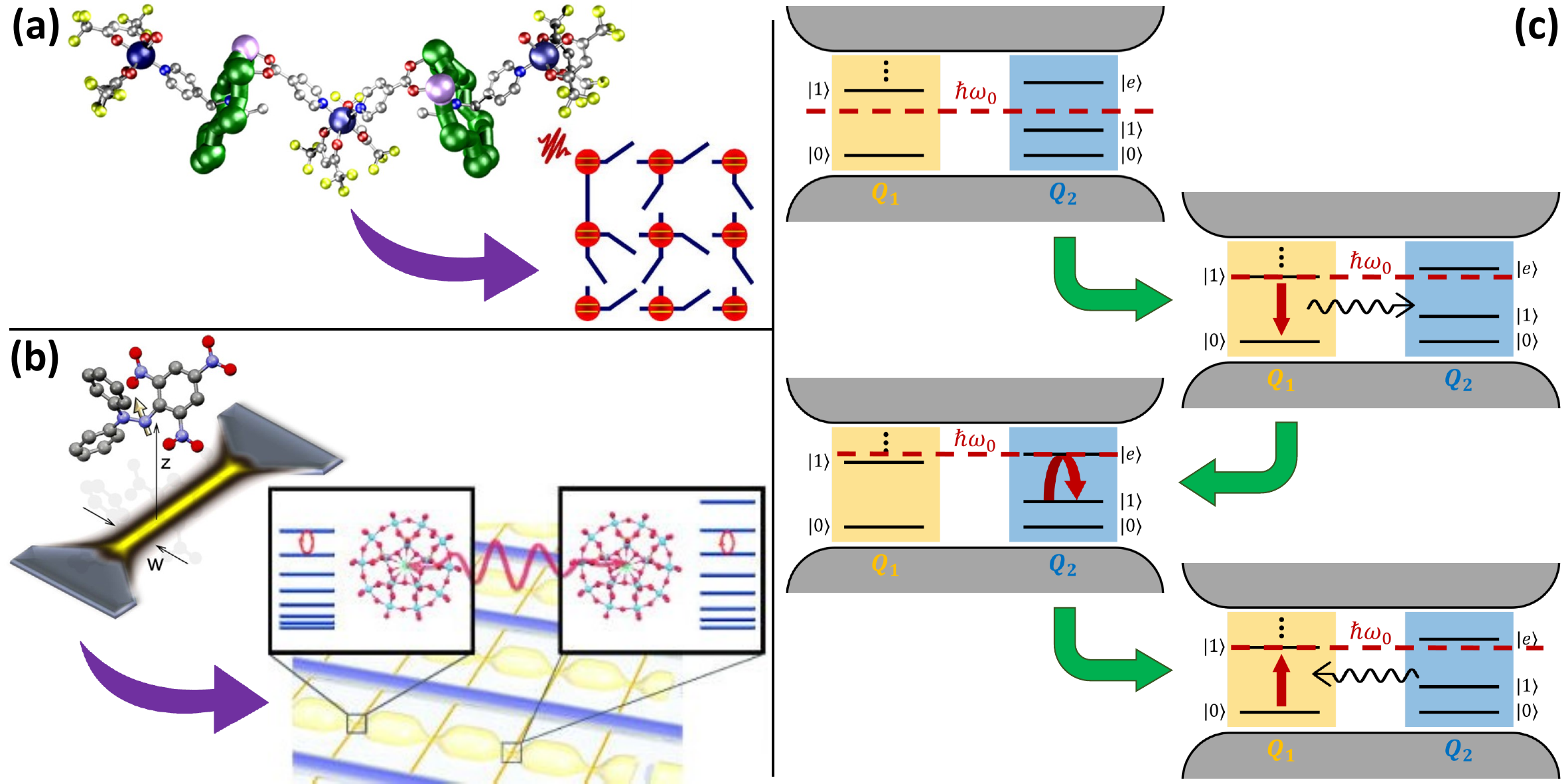}
	\caption{Scalability of the molecular spin quantum processor by exploiting (a) the capabilities of Chemistry to combine several spins into complex supra-molecular structures \cite{Lockyer2022} and/or (b) the possibility to deposit molecules within superconducting resonators \cite{Gimeno2020,Carretta2021}, whose layout can be designed in order to maximize the magnetic field in proper constrictions. In (a) the molecule is a chain of 5 qubits: three Cu$^{2+}$ complexes (blue spheres) and two Cr$_7$Ni rings (green octagons) \cite{Lockyer2022}. (c) Scheme for the implementation of a controlled-phase gate between two molecular spin qudits, strongly coupled to the photon field within a superconducting resonator. The photon is brought in and out of resonance with different spin transitions and mediates an effective interaction between the two spins. In the idle phase (i) the resonator is empty (no photons) and the photon frequency is very different from all the spin energy gaps and hence the two spins are decoupled. Then the photon frequency is tuned into resonance with the $\ket{1}-\ket{0}$ transition of $Q_1$, yielding a resonant photon emission (ii). The photon frequency is then modified to match the $\ket{1}-\ket{e}$ energy gap of the second qudit and a semi-resonant $2\pi$ oscillation is implemented (iii), thus adding a phase to the $\ket{11}$ component of the two-qudit state. Finally, the photon re-emitted in (iii) is absorbed to induce again the $\ket{0}-\ket{1}$ transition, restoring the initial populations, with the phase added in step (iii). 
    Panel (a) reprinted with permission from \cite{Lockyer2021}. Panel (b) reprinted with permission from \cite{Carretta2021}).}
	\label{fig:scalability}
\end{figure*}

The multi-qubit structures presented above can be  extended to include several interacting spin qubits. An example of a 5-qubit chain is shown in Fig. \ref{fig:scalability}-(a), consisting of alternating Cu$^{2+}$ and Cr$_7$Ni complexes with a tailored hierarchy of interactions which can be exploited for specific quantum simulation algorithms \cite{Lockyer2022}.
In spite of the impressive degree of chemical control, these permanently-coupled multi-qubit structures in general suffer from residual interactions which are never completely switched off and hence limit this "chemical" scalability to less then $\sim 10$ qubits \cite{modules}. \\
Indeed, a fully-scalable architecture requires the capability  to switch on inter-qubit couplings only during the implementation of two-qubit gates and to switch them off completely (or at least to the largest extent) in the other computational steps. Any residual interaction in these stages (the so-called {\it cross-talk}) will introduce errors in the implemented algorithm. Although unitary, this dynamics is hardly correctable in a deterministic way when the size of the hardware increases. To achieve this, other methods besides sythetic tools can be considered to design a multi-qubit register based on molecular spins.

An alternative promising route is borrowed from the technology largely developed for the control of superconducting qubits, based on superconducting resonators [Fig. \ref{fig:scalability}-(b)] \cite{Jenkins2016,Chiesa2023prappl}. A resonator can host several non-interacting molecular spin qubits or qudits, strongly coupled to the quantized photon field. 
Let us consider for instance a couple of spins $S \ge 1/2$ within a superconducting resonator. 
These can be described by the Hamiltonian:
\begin{eqnarray} \nonumber
	H_{sp} &=& \hbar \omega_0(t) a^\dagger a + \mu_B B \sum_i g_i s_{zi} + D_i S_{zi}^2 \\
	&+& \sum_i 2 G_i S_{xi} \left( a + a^\dagger \right) 
	\label{eq:dicke}
\end{eqnarray}
where $a^\dagger$ ($a$) are creation (annihilation) operators of the photon field of energy $\hbar \omega_0$. $S_{\alpha i}$ are spin $\ge 1/2$ operators whose energy gaps are made different by the combination of Zeeman and zero-field splitting ($D_i S_{zi}^2$) terms, typical of this kind of systems.
Different spins will also be in general inequivalent, i.e. characterized by different values of $D_i$ and $g_i$.
Finally, $G_i$ is the coupling between individual spins and photons. Note that the frequency of the resonator $\omega_0(t)/2\pi$ can be tuned, e.g. by proper SQUID devices \cite{Palacios-Laloy2008}.
In the idle stage, the photon frequency is largely detuned from the spin gaps (of an amount $\gg G_i$) and hence the spin-photon interaction appearing in Eq. \eqref{eq:dicke} is completely ineffective. 
However, it can be exploited to mediate an effective coupling between different qubits via a sequence of resonant photon absorptions and emissions \cite{Chiesa2023prappl}. We now  define a computational space with no photons in the resonator and the states of the qubit pair encoded into the two lowest energy levels of the two spins [see Fig. \ref{fig:scalability}-(c), top panel]. 
By tuning the resonator frequency into resonance with the spin gap of the first qubit, a photon is emitted [Fig. \ref{fig:scalability}-(c), second panel]. 
As a second step, by further tuning of the resonator, the photon frequency is brought into resonance with the gap between the first and second excited state of the second qubit. By semi-resonant absorption and re-emission, its state is ultimately brought back to $\ket{1}$ with an additional phase (third panel). Further absorption of the photon by the first spin (with proper cavity tuning) brings back to the original $\ket{11}$ state, apart from a phase (fourth panel). Hence, we have exploited the second excited state of the second spin as an auxiliary state to implement a controlled-phase operation. Conversely, other computational states are unaffected because the addressed transitions are either off-resonance or do not occur because the are no photons available.

A few comments are in order. First, an array of resonators can be realized, with photons jumping between them each time a pair of neighboring resonators is tuned into mutual resonance \cite{Carretta2013,Chiesa2023prappl}. Hence, this architecture is able to establish long-range entanglement between distant molecular qudits by exploiting the natural mobility of photons. Second, the scheme illustrated above for qubits can be applied to any two-{\it qudit} state of a given pair of molecular spins $>1/2$. This enables the implementation of generic two-qudit antangling gates and hence settles the basis for multi-qudit logic.  \\
Finally, we note an important difference with respect to superconducting qubits. The latter, indeed, are artificial atoms coupled to the electric (and not to the magnetic) component of the electromagnetic field. Here instead, the single-spin to photon coupling is many orders of magnitude weaker, because it occurs via Zeeman interaction with the magnetic field component. Hence, one of the greatest challenges for transforming  molecular spin qubits into an actually working quantum computing platform is represented by achieving the strong coupling between individual spins and cavity photons, i.e. a regime in which the coupling overcomes the decoherence rates of the system and thus allows for coherent manipulations. To this end, the idea is to combine proper engineering of the resonators layout and of the molecular spin system. For instance, realising nm-size constrictions where the electro-magnetic field is confined and strongly concentrated enabled to enhance $G_i$ by more than 3 orders of magnitude \cite{Gimeno2020}. 
Moreover, specific transitions in spin systems with large $S$ can be exploited to increase the effective coupling \cite{Chiesa2023prappl}. Indeed, the matrix element for a $\ket{m}\rightarrow \ket{m\pm 1}$ transition is given by
\begin{equation}
   \bra{m} S_x \ket{m\pm 1} = \frac{1}{2} \sqrt{S(S+1) - m (m\pm 1)} ,
    \label{eq:Smatel}
\end{equation}
which is maximum if we start from state $\ket{m=0}$, i.e. from the ground state of a MNM with the so-called {\it easy-plane} anisotropy ($D_i>0$ in Eq. \eqref{eq:dicke}) or if we consider spin-clock transitions occurring between superpositions of $(\ket{S} \pm \ket{-S})/\sqrt{2}$ states \cite{Gimeno2023} (see Sec. \ref{sec:decoqudit} below).
In these conditions, a factor of $\sim S$ in the spin-photon coupling can be gained. This
value can be significant if we consider, e.g., $S=10$ (i.e. the ground total spin state of famous multi-spin clusters such as Fe$_8$ or Mn$_{12}$).

\section{Decoherence}
\label{sec:deco}

\subsection{General concepts}
Quantum information encoded into qubits is corrupted by the interaction between qubits and environment, a phenomenon which is generally called decoherence. \\
A proper description of decoherence requires the introduction of the density operator $\rho$. This allows us to treat situations in which we do not exactly know the system state-vector, but only the probabilities $p_n$ of finding the system in state-vector $\ket{\psi_n}$.
In this case, we introduce the density operator
\begin{equation}
	\rho = \sum_n p_n \ket{\psi_n} \bra{\psi_n}.
	\label{eq:rho}
\end{equation}
We call this generic state of the system a {\it mixed} state, while if $p_n = \delta_{nk}$ (i.e. we are sure to find the system in $\ket{\psi_k}$) we recover a {\it pure} state and the description of Quantum Mechanics in terms of the density operator is equivalent to that based on the state-vector.
However, the density-matrix formulation is more general.
In particular, it is very useful to describe the dynamics of a subsystem which is part of a larger quantum system. For instance, we could be interested in measuring a local observable acting only on one or a few qubits in the register. Moreover, the interaction of the qubits in a quantum hardware with a large number of environmental degrees of freedom can be seen as the evolution of a small {\it open} subsystem of interest which is part of a larger Hilbert space.

Before introducing a formal description of the dynamics of an open quantum system, let us consider a simple illustrative example on a pair of qubits to understand the effect of decoherence on a quantum state.
In this example, the first qubit models the system and is labeled by S, the second represents the environment (E).
S is initialized in a generic pure state $\ket{\psi}=\alpha \ket{0}+\beta \ket{1}$, while E in $\ket{0}$.
Hence, the whole S+E is initially in a factorized state $\ket{\Psi} = \ket{\psi} \otimes \ket{0}$.
Then, a conditional rotation is applied to E, depending on the state of S, i.e.
\begin{eqnarray} 
	\ket{0} \otimes \ket{0} &\rightarrow& \ket{0} \otimes \ket{0} \\  \nonumber
	\ket{1} \otimes \ket{0} &\rightarrow& \ket{1} \otimes \left( \sqrt{1-p} \ket{0} + \sqrt{p} \ket{1} \right) ,
	\label{eq:decodid}
\end{eqnarray}
with $p$ a real number such that $p \le 1$.
We now focus only on the state of S, by taking the {\it partial trace} over E. This allows us to derive the final S density matrix, regardless from the state of E.
Formally, we need to compute $\rho_S = \bra{0_E} \rho \ket{0_E} + \bra{1_E} \rho \ket{1_E} $, where $\rho= \ket{\Psi} \bra{\Psi}$ is the whole S+E density matrix. Hence, we get
\begin{eqnarray} \nonumber
	\rho_S &=& |\alpha|^2 \ket{0} \bra{0} + \alpha \beta^* \sqrt{1-p} \ket{0} \bra{1} \\
	   &+& \alpha^* \beta \sqrt{1-p} \ket{1} \bra{0} + |\beta|^2 \ket{1} \bra{1},
    \label{eq:decodid2}
\end{eqnarray}
i.e. in matrix form
\begin{equation}
	\rho_S = 
	\begin{pmatrix}
		|\alpha|^2 & \alpha\beta^*\sqrt{1-p} \\
		\alpha^*\beta\sqrt{1-p} & |\beta|^2
	\end{pmatrix} .
\end{equation}
If we compare this expression with that of a pure state $\rho_S^{(p)} = \ket{\psi} \bra{\psi}$ we note that the diagonal elements (populations) are the same, while off-diagonal ones (coherences) are decreased by a factor $\sqrt{1-p}$.
Hence, entanglement between S and E, followed by partial trace over E has introduced an effect usually known as {\it pure dephasing}, in which coherences in the S eigenbasis are lost by an amount which depends on the S-E interaction (here parameterized by $p$). In the $p=1$ case coherences vanish and the final state is a {\it classical} mixture of $\ket{0}$ and $\ket{1}$. \\
This can be extended to what happens when an open quantum system interacts with the environment. First, the system-environment coupling can introduce entanglement between the two. Then, information on the bath is lost due to the large number of its degrees of freedom. This corresponds to tracing out the bath and reducing the originally pure and very large density matrix to a mixed one on a subset of degrees of freedom.  \\
Remarkably, as in the example of Eq. \eqref{eq:psimmix} (Sec. \ref{sec:2qubits}), the resulting evolution is {\it non-unitary}, i.e. there is no pure state $\ket{\phi}$ such that $\rho_S=\ket{\phi}\bra{\phi}$. $\rho_S$ is a mixture and it cannot arise from a Schr\"odinger equation acting on a state-vector. 
Conversely, the general density-matrix picture is able to model both unitary and non-unitary dynamics.

In general, an accurate description of the bath can be difficult and is strongly system-dependent. By introducing some approximations \cite{Breuer}, we can write the following Lindblad equation for $\rho_S$:
\begin{equation}
	\dot{\rho}_S = -\frac{i}{\hbar} [H,\rho_S] + \gamma \left( 2 x \rho_S x^\dagger  - x^\dagger x \rho_S - \rho_S x^\dagger x \right) ,
\label{eq:lindblad}
\end{equation} 
where the first term represents the coherent (unitary) evolution, while the second describes the non-unitary effect of decoherence, driven by operator $x$ at a rate $\gamma$.
The underlying assumptions (see, e.g. \cite{Breuer} for a detailed derivation) are the following: a weak system-environment interaction (Born approximation), no-memory in the bath or very short bath correlation times (Markov) and independent evolution of diagonal and off-diagonal elements of $\rho$ (secular approximation).
The Lindblad equation can be derived both starting from a microscopic description of the system-environment interaction (from which the operators $x$ acting on the system are deduced) or obtained from a mathematical construction which ensures general properties of $\rho$, such as complete positivity. In this latter case, the operator $x$ can be a generic guess of the effect of noise which corrupts the coherent evolution of the system.

In QIP, an alternative useful description is given by the operator-sum representation of $\rho_S$ \cite{Nielsen}.
This is an expansion of the system density matrix at  time $t$ in terms of {\it Kraus} operators $E_k$ acting on the initial $\rho_S(0)$, i.e.:
\begin{equation}
	\rho_S(t) = \sum_k E_k \rho_S(0) E_k^\dagger.
\end{equation}
Depending on the specific form of $E_k$ operators, this expression can include both coherent and incoherent evolution.
This second approach is rather different from above. Instead of solving Eq. \eqref{eq:lindblad} and follow the continuous time-evolution of $\rho_S$, we only consider the dynamics at discrete time-steps. \\
When analyzing the output of a quantum circuit, this is very practical: indeed, we can introduce noise simply by adding a proper operator after each quantum gate or idle step in the circuit. This allows one to obtain an effective description of noise and of its effect on a quantum algorithm regardless of its specific microscopic origin. 
For instance, a common choice is that of introducing a "depolarizing channel" to model noise affecting quantum operations. For a single qubit, this corresponds to assuming $\sigma_x$, $\sigma_y$ or $\sigma_z$ error operators $E_k$ acting with the same probability on the qubit state \cite{Nielsen}. The same idea can be generalized to multi-qubit gates and operations. 
The drawback of this approach is that these noise-models can represent a rather crude approximation of reality and hence hinder a proper description of the dynamics, as well as the design of targeted strategies to mitigate errors. 

Therefore, we focus in the following on hardware-related noise models. In particular, in many solid-state platforms the two most important decoherence channels are given by pure dephasing and relaxation.
We will discuss the origin of both phenomena in molecular spin systems in the next subsection. 
Here we only note that in the basis of the Hamiltonian eigenstates, the former only induces a decay of the coherences in the system density matrix (see above example), while the latter affects both diagonal and off-diagonal elements.
This can be easily seen on a single qubit, both in the Lindblad and operator-sum descriptions. \\
Pure dephasing is represented by the diagonal Lindblad operator $x=s_z$ or equivalently by Kraus operators $E_0 = \sqrt{(1+p)/2} I$ and $E_1 = \sqrt{(1-p)/2} \sigma_z$. Indeed, neglecting the coherent evolution, Eq. \eqref{eq:lindblad} can be easily solved, getting $\bra{0} \rho_S (t) \ket{1} = \bra{0} \rho_S (0) \ket{1} e^{-\gamma t}$, while diagonal elements of $\rho$ are not affected. 
The characteristic decay time of the coherence is usually called coherence time $T_2 = 1/\gamma$. The same result is obtained by the operator-sum representation, after having set $p=e^{-t/T_2}$. \\
The effect of relaxation in general depends on the temperature: population is transferred between system eigenstates in order to approach thermal equilibrium, i.e. a diagonal density matrix with populations corresponding to Boltzamnn statistics $\rho_S = \sum_n e^{-E_n/k_BT}/Z \ket{\psi_n} \bra{\psi_n}$, with $E_n$ and $\ket{\psi_n}$ the system eigenvalues and eigenstates, $T$ the temperature and $Z$ the partition function. In multi-level MNMs, the relaxation process can be very complex and involve several characteristic times and regimes. This has been the subject of many studies over the years.
The situation is much simpler for a single qubit at $T\rightarrow 0$. In this case thermal equilibrium is represented by $\rho_S = \ket{1}\bra{1}$, i.e. relaxation only leads to population in the ground state. The corresponding evolution of $\rho_S$ is obtained from Lindbald description by setting $x=s_-$ and $\gamma = 1/2T_1$, with $T_1$ the characteristic relaxation time. 
Alternatively, the same result arises by choosing Kraus operators $E_0 = \ket{1} \bra{1} + \sqrt{e^{-t/T_1}} \ket{0}\bra{0}$ and $E_1 = \sqrt{1-e^{-t/T_1}} \ket{1}\bra{0}$.
Note that besides the decay of the excited state population with a rate $1/T_1$, we get also a relaxation-induced dephasing with a halved rate $1/2T_1$.

\subsection{Molecular Nanomagnets: spin Hamiltonian and main decoherence mechanisms}
\label{sec:decoqudit}

We now move to a closer description of the molecular spin quantum hardware and to the main decoherence mechanisms characterising it. To this end, it is useful to introduce the typical spin Hamiltonian describing molecular spin qudits \cite{BOOK}.
We consider, in particular, two classes of MNMs: either (i) strongly-interacting multi-spin clusters consisting of transition metal ions, or (ii) nuclear spin qudits consisting of a single transition metal or rare-earth ion with an electronic spin coupled by a strong hyperfine interaction to a nuclear spin $I\ge 3/2$. 

In the first case (i) the most important contributions to the spin Hamiltonian are the following:
\begin{equation}
    H_{3d} = H_{iso} + H_{Z} + H_{zfs} + H_{an}.
\end{equation}
Here the isotropic exchange is the leading term:
\begin{equation}
    H_{iso} = \sum_{ij} J_{ij} {\bf s}_i \cdot {\bf s}_j .
\end{equation}
This interactions is spherically symmetric and commutes with $S^2$, ${\bf S} = \sum_i {\bf s}_i$ being the total spin of the molecule. Therefore, the spectrum of $H_{iso}$ consists of $(2S+1)-$degenerate multiplets characterised by a well defined total spin $S$. These multiplets are then split by the Zeeman interaction 
\begin{equation}
    H_Z = \mu_B \sum_i {\bf s}_i \cdot {\bf g} \cdot {\bf B} , 
\end{equation}
where ${\bf B}$ is the external field and ${\bf g}_i$ the spectroscopic tensor of ion $i$. Additional anisotropic terms $H_{zfs}$ and $ H_{an}$ usually act perturbatively and cause  a slight mixing between states belonging to different multiplets. 
In particular, the zero-field splitting is a single-ion term of the form
\begin{equation}
    H_{zfs} = \sum_i \sum_{k\le2s} \sum_{q=-k}^k b_k^q O_k^q(s_i) , 
    \label{eq:zfs}
\end{equation}
where $O_k^q$ are Stevens equivalent operators and $k$ is even. 
The most common contributions to $H_{zfs}$ are the rank $k=2$ axial $d_i s_{zi}^2$ and rhombic $e_i (s_{xi}^2-s_{yi}^2)$ terms. They are only effective on spins $s_i \ge 1$ qudits, with the important effect of making the spectrum {\it anharmonic}, i.e. to distinguish the gaps of different transitions between the eigenstates (as already mentioned in Sec. \ref{sec:2qubits}.B). \\
Finally, $H_{an}$ includes anisotropic and anti-symmetric contributions to the spin-spin couplings:
\begin{equation}
\label{eq:Haniso}
    H_{an} = \sum_{ij} {\bf s}_i \cdot {\bf D}_{ij} \cdot {\bf s}_j  + \sum_{ij} {\bf G}_{ij} \cdot {\bf s}_i \cross {\bf s}_j,
\end{equation}
where ${\bf D}_{ij}$ and ${\bf G}_{ij}$ are the anisotropic exchange tensor (including also dipolar couplings) and anti-symmetric (Dzyaloshinskii-Moriya) vector, respectively. The properties of anisotropic and anti-symmetric exchange are rather different (they are represented by tensor operators of rank 2 and 1, respectively), but we resend to specialised books for details \cite{BOOK}.

Molecular nuclear spin qudits (ii) are well described by a spin Hamiltonian of the form
\begin{equation}
\label{eq:nuclei}
H_{mnq} = {\bf I} \cdot {\bf A} \cdot {\bf s} + p I_z^2 +  \mu_B \; \textbf{s} \cdot \textbf{g} \cdot \textbf{B},
\end{equation}
where $I$ is the nuclear spin, ${\bf A}$ is the hyperfine coupling tensor and $s$ is the electronic spin. We focus in particular on electronic spin doublets, but interesting examples of coupled electronic and nuclear qudits also exist (consider for instance compounds containing $^{55}$Mn$^{2+}$, where $I=s=5/2$).
The second term in Eq. \eqref{eq:nuclei} is a nuclear quadrupole interaction (similar to $H_{zfs}$ but considered only axial for simplicity), while the last is the Zeeman coupling of the electronic spin with the external field. We have neglected the very weak nuclear Zeeman interaction.
In the context of QIP, we usually consider a regime in which the Zeeman interaction is the leading term in $H_{mnq}$ and hence the quantisation axis is set by the direction of ${\bf B}$ for both electronic and nuclear spins (due to the large hyperfine).
The corresponding eigenstates of $H_{mnq}$ are practically factorised into a nuclear and an electronic spin component $\ket{m_I} \otimes \ket{m_s}$ along ${\bf B}$. In transition metal and rare-earth ions, $A$ ranges from hundreds of MHz to $\sim 1$ GHz. Hence, the component of ${\bf A}$ parallel to ${\bf B}$ tend to align electronic and nuclear spins, while 
the components of ${\bf A}$ perpendicular to ${\bf B}$ induce a mixing between electronic and nuclear spins. Although this effect is usually small on the eigenvectors ($A_\perp/g\mu_B B \ll 1$), it can be very important for manipulating the state of the nuclear qudit. Indeed, thanks to this small mixing, nuclear spin transitions become much faster \cite{VOTPP,jacsYb} and different energy gaps are well resolved, even when $p = 0$. \\

After this brief introduction of the physical system, we are now in a position to describe dephasing and relaxation in MNMs.
In particular, at not too-low temperatures, relaxation is driven by the coupling of the system spins with vibrational modes. The slow relaxation of a particular class of MNMs (the so called single-molecule magnets) has been and is still nowadays the subject of intense research oriented to increase the operational temperatures of these systems, where a single {\it classical bit} of information could be stored and kept in a single molecule, thus reaching an unprecedented storage density \cite{Nat_dycene,Sci_dycene,prbDycene,Garlatti2021}.
The development of better devices proceeds here along two parallel lines, based on engineering on the one hand the spin Hamiltonian and on the other molecular vibrations, i.e. both the phonon spectrum and spin-phonon couplings. These couplings arise from the modulation of different terms in the spin Hamiltonian, induced by atomic displacements of different forms and energies \cite{Garlatti2023,Lunghi2017}. \\ 
Understanding the phonon-driven relaxation dynamics is also important to develop molecular qubits in which this effect becomes negligible. Many theoretical studies on VO complexes have been carried on to unravel the role of different kinds of phonons at different energy scales, in some cases increasing the operating temperature of molecular spins \cite{Atzori2017,Albino2019,TesiDalton2016}.
Note that $T_1$ also sets a upper limit of $2T_1$ on the coherence time $T_2$, i.e. the characteristic decay time of the off-diagonal elements of $\rho_S$ (see end of Sec. \ref{sec:deco}.A). Moreover,  it has been recently shown that phonons can also introduce an additional pure dephasing channel via Raman mechanisms  \cite{Mondal2022}. \\
At low temperatures, where most quantum computing platforms operate, phonon-induced decoherence mechanisms are largely suppressed and pure-dephasing becomes completely dominant. Indeed, both phonon absorption and phonon emission are very unlikely. The former because thermal population of the excited states is negligible (molecular qubits' energy gaps are much larger than $k_BT$), the latter because the probability of phonon emission scales as the third power of the energy gap \cite{Wurger1998,Santini2005}, which is significantly smaller than the Debye energy in these compounds. In practice, many experimental evidences exist that at low temperatures $T_1$ becomes very long and hence it does not limit the system coherence (see, e.g., \cite{SIMqubit,Bader2016}). 

\begin{figure}
	\includegraphics[width=0.48\textwidth]{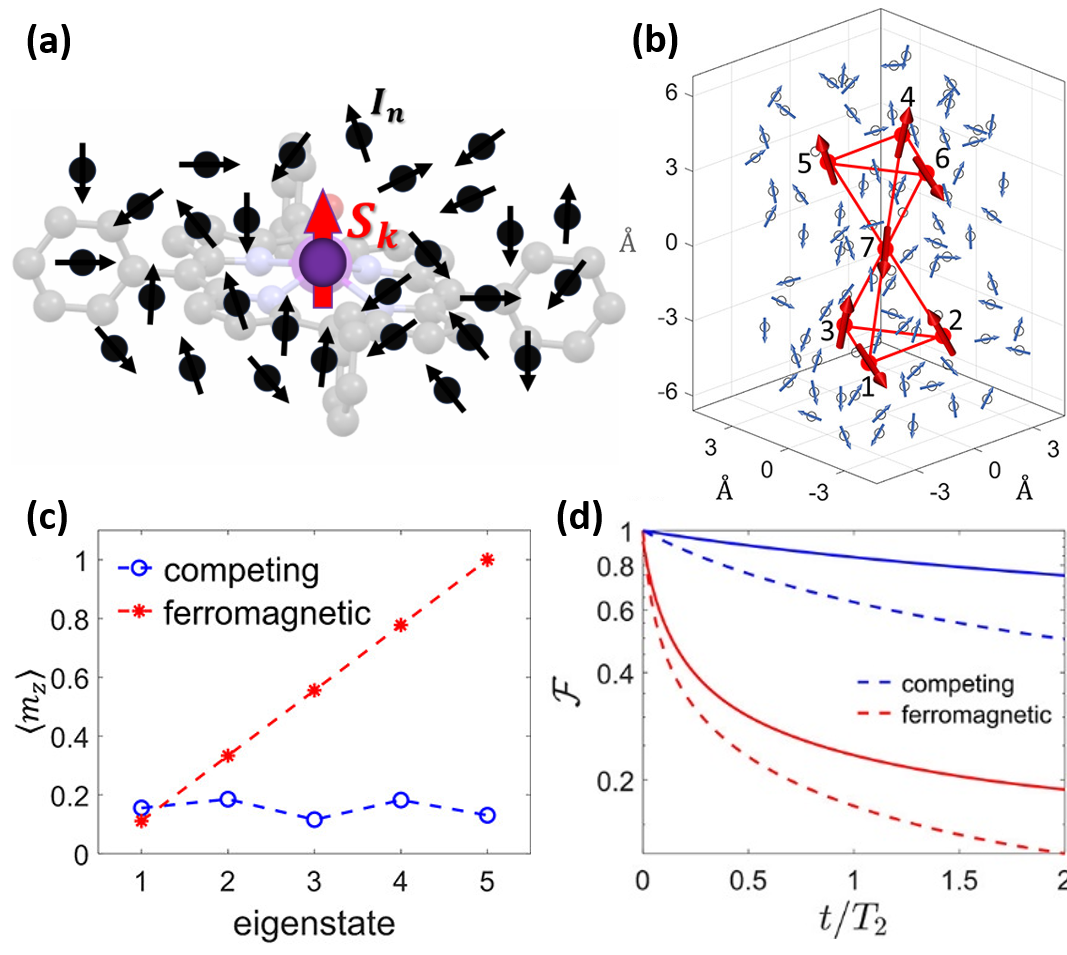}
	\caption{(a) A schematic representation of the nuclear spin bath (small black arrows) interacting with the 
        central molecular spin (big red arrow) according to Eq. \eqref{eq:dipolar}. On the background the structure of a vanadyl phthalocyanine molecular qubit \cite{Atzori_JACS} is reported. (b) Double-tetrahedron structure of a 7-ion molecule (such as Ni$_7$ \cite{Ni7}) surrounded by a nuclear spin bath. (c) Calculated values of $\braket{m_z}=\sum_{i=1}^7\braket{s_i^z}/s_i$ on the lower-energy eigenstates of the system in panel (b) for both ferromagnetic and competing antiferromagnetic interactions. (d) Fidelity over time of a generic superposition (dashed lines) or an encoded state (solid lines, see Section \ref{sec:qec}) for both ferromagnetic and competing antiferromagnetic interactions. Fidelity is calculated between the initial state and the state at time $t$ for the system in panel (b). Panels (b), (c) and (d) reprinted with permission from [\onlinecite{Chiesa2022}].}
	\label{fig:spin_bath}
\end{figure}

Hence, increasing $T_2$ represents the most important focus for the design and synthesis of well-working molecular spin qubits. In isolated MNMs at low temperature, pure dephasing originates from the interaction between the central spins and the nuclear spins belonging to the ligands surrounding them within each molecule \cite{npjQI}, as shown schematically in Figure \ref{fig:spin_bath}-(a). 
Note that spin-spin dipolar interactions between different molecules in a crystal or in solution can play a role, but this contribution is also expected to reduce with temperature \cite{Ghirri2015} and it is not relevant if we focus on isolated molecules or in molecules diluted in a diamagnetic crystal \cite{Ga7Zn}. 

The interaction between central spins [a single spin $S$ in Fig. \ref{fig:spin_bath}-(a)] and the nuclear spin bath 
is of the form $D_{kn}^{\alpha\beta} S_k^\alpha I_n^\beta$, where $S_k^\alpha$ is the $\alpha$ spin component of the central spin $k$ within the molecule, $I_n^\beta$ is the $\beta$ component of nuclear spin $n$ and the coupling tensor has the form:
\begin{equation}\label{eq:dipolar}
    D_{kn}^{\alpha\beta} =  g_kg_N\mu_B\mu_N(3r_{kn}^\alpha r_{kn}^\beta/\abs{r_{kn}}^2 - \delta_{\alpha\beta})/\abs{r_{kn}}^3
\end{equation}
where $g_k$ is the central spin spectroscopic factor, $g_N$ the nuclear spin spectroscopic factor (both assumed isotropic for simplicity), $\mu_N$ is the nuclear magneton and $r_{kn}^\alpha$ is the $\alpha$ component of the distance between the central spin $k$ and the $n$-th nuclear spin.
In this expression, off-diagonal interaction terms $\alpha = x,y$ are practically ineffective, because 
the electronic energy gap of the central spin is much larger than $D^{\alpha\beta}_{kn}$. Conversely, diagonal terms which do not induce a flip of the central spin ($\alpha = z$) are effective and responsible for decoherence. Indeed, they induce a dynamics on the bath which is different depending on the state of the central spin. If the spins in the bath are static, i.e. if we neglect the interactions between them, this evolution can be exactly canceled by spin-echo pulse sequences.  \\
Conversely, an evolving bath (similarly to a fluctuating magnetic field) induces an irreversible decay of the coherence in the system density matrix, exactly as described in Sec. \ref{sec:deco}.A. Such a decay [$\sqrt{1-p}$ in Eqs. (\ref{eq:decodid}-\ref{eq:decodid2})] is directly associated to the overlap between the  many-body state of the bath for the  different states of the central spins \cite{Troiani2008}: it is maximum when system and bath are 
strongly entangled [$p \rightarrow 1$ in Eq. \eqref{eq:decodid}] and it disappears when they evolve independently ($p \rightarrow 0$). 
The exact calculation of how coherences decay would require a quantum computer, but several techniques exist to obtain good approximations of this complex time evolution (such as the cluster-correlation expansion method \cite{npjQI}). 

The coherence time of molecular spin qubits and qudits can be engineered by working both on the system Hamiltonian and on the nuclear spin bath. On the synthetic side, several efforts have been devoted to realize structures with very few nuclear spins, achieving coherence times in the ms range \cite{Zadrozny2015}. On the theoretical side, specific encodings can be found based on multi-spin clusters which are substantially protected from decoherence.
Among these, we recall the use of the chirality degree of freedom in spin triangles \cite{Troiani2008,Troiani2012}, which is not subject to first-order decoherence mechanisms. \\
An important coherence enhancement can also be achieved by considering logical states $\ket{0}$ and $\ket{1}$ encoded in symmetric and anti-symmetric superpositions of $\ket{S}$ and $\ket{-S}$ states of a spin $S$ \cite{Hill,Santamarina2020,Gaita2022}, i.e. the so-called clock-transition qubits. This situation can be obtained by choosing specific magnetic fields where the system exhibits level avoided crossings, due to transverse anisotropy terms in the spin Hamiltonian. For instance, for a spin $S=1$ at high magnetic field (parallel to the easy anisotropy axis), $\ket{m=\pm S}$ are eigenstates of the system. If however the magnetic field is tuned to make them approximately degenerate, a transverse anisotropy term $E(S_x^2-S_y^2)$ will mix them into $(\ket{S} \pm \ket{-S})/\sqrt{2}$, with an energy gap $2E$ at the avoided crossing. These states encode a qubit substantially protected from magnetic field noise, because at the avoided crossing the gap (and hence the transition energy) is insensitive to small magnetic field changes. As a result, the coherence time of the corresponding superposition of logical states is enhanced \cite{Hill}. \\

In multi-spin molecules consisting of several interacting $3d$  spins, it can be shown that (in the Lindblad approximation) coherences of $\rho_S$ show a mono-exponential decay. The decay rate associated with a superposition between eigenstates $\ket{\mu}$ and $\ket{\nu}$ takes the form:
\begin{align}\label{eq:gamma}
     \gamma_{\mu\nu} =& \sum_{j j^\prime} C_{jj^\prime} [    
     \langle \mu | s_j^z | \mu \rangle \langle \mu | s_{j^\prime}^z | \mu \rangle   \nonumber\\ 
     +& \langle \nu | s_j^z | \nu \rangle \langle \nu | s_{j^\prime}^z | \nu   \rangle - 2 \langle \mu | s_j^z | \mu \rangle \langle \nu | s_{j^\prime}^z | \nu \rangle ].
\end{align}
Here $C_{jj^\prime}= \sum_{nn^\prime} \sum_{\beta\beta^\prime} \chi_{nn^\prime}^{\beta\beta^\prime}(0) D_{jn}^{z\beta} D_{j^\prime n^\prime}^{z \beta^\prime}$ is a coefficient related to sum of products of dipolar interactions between the central spin and the nuclear spin bath \cite{Chizzini2022} and to the zero-energy bath spectral function $\chi_{nn^\prime}^{\beta\beta^\prime}(0)$. 
A part from this term dependent on the specific molecular structure, a key piece of information is contained in the expectation values of local spin operators $\bra{\mu} s_j^z \ket{\mu}$, which are only related to the eigenstates and to some general properties of them. 
In particular, it can be shown that if different spins are coupled by anti-ferromagnetic competing exchange interactions a particularly favorable pattern of $\bra{\mu} s_j^z \ket{\mu}$ \cite{Chizzini2022,Chiesa2022} arises, which strongly suppresses decoherence.  
Indeed, anti-ferromagnetic (isotropic) couplings between pairs of spins in the molecule will lead to low-energy eigenstates characterized by low total spin and in general small values of $\bra{\mu} s_j^z \ket{\mu}$. The situation is particularly advantageous if compared to an opposite one, in which the spins subject to the same nuclear spin bath are all ferro-magnetically coupled. In that case, the ground state multiplet is characterized by a large total spin $S = \sum_j s_j$ and (assuming for simplicity an axial Hamiltonian with a magnetic field along $z$) the eigenstates are the same of $S_z = \sum_j s^z_j$. Hence, the low-energy spectrum is equivalent to that of a single spin $S$ ion. 
A prototypical example of this behavior is reported in Fig. \ref{fig:spin_bath}-(b-d), where (following Refs. \cite{Chiesa2022,Chizzini2022}) we consider a 7-ion cluster with the same structure of Ni$_7$ molecule \cite{Ni7} and we compare
the two situations with either ferromagnetic or competing exchange interactions. In the former case, the values of $\bra{\mu} s_j^z \ket{\mu}$ vary significantly with $\ket{\mu}$, since different eigenstates are magnetically very different. Conversely, with competing spin-spin interactions the variation of $\bra{\mu} s_j^z \ket{\mu}$ is much smaller. As a consequence, a superposition of all the qudit states is much more protected from decoherence if the hierarchy of interactions in the molecule is chosen properly (see Fig. \ref{fig:spin_bath}-(d)). This can have very important implications in algorithms explicitly exploiting the molecular multi-level structure, i.e. where the elementary unit of information is a qudit instead of a qubit \cite{Chizzini2022}. Especially, it can be exploited to encode a protected qubit of information very efficiently into a single qudit. This idea of embedding a protected qubit into a single object (and not into a collection of different qubits as commonly proposed) is compatible with chemical structure engineering and could greatly simplify the actual implementation of quantum error correction (QEC). As discussed below, achieving quantum error correction is a mandatory step for any quantum computing platform to become really useful.

\subsection{Embedding quantum-error correction in molecular spin qudits}\label{sec:qec}
 

Decoherence processes described above are extremely harmful for quantum computers, which thrive on fragile quantum superpositions which are disrupted by these effects. 
Hence, proper strategies to suppress errors must be identified and this is why QEC is crucial. The ultimate goal is not only to preserve quantum information, but to achieve {\it fault-tolerant} quantum computation, i.e. the capability of performing an arbitrary number of quantum gates by always keeping the error small, even on a faulty hardware. This is a very demanding task, and part of this is because a classical {\it majority voting} approach (commonly used in preventing errors in classical bits) would require copying the information of a qubit on another one, which is forbidden by the {\it no cloning theorem} \cite{Nielsen}. 

\begin{figure}
	\includegraphics[width=0.48\textwidth]{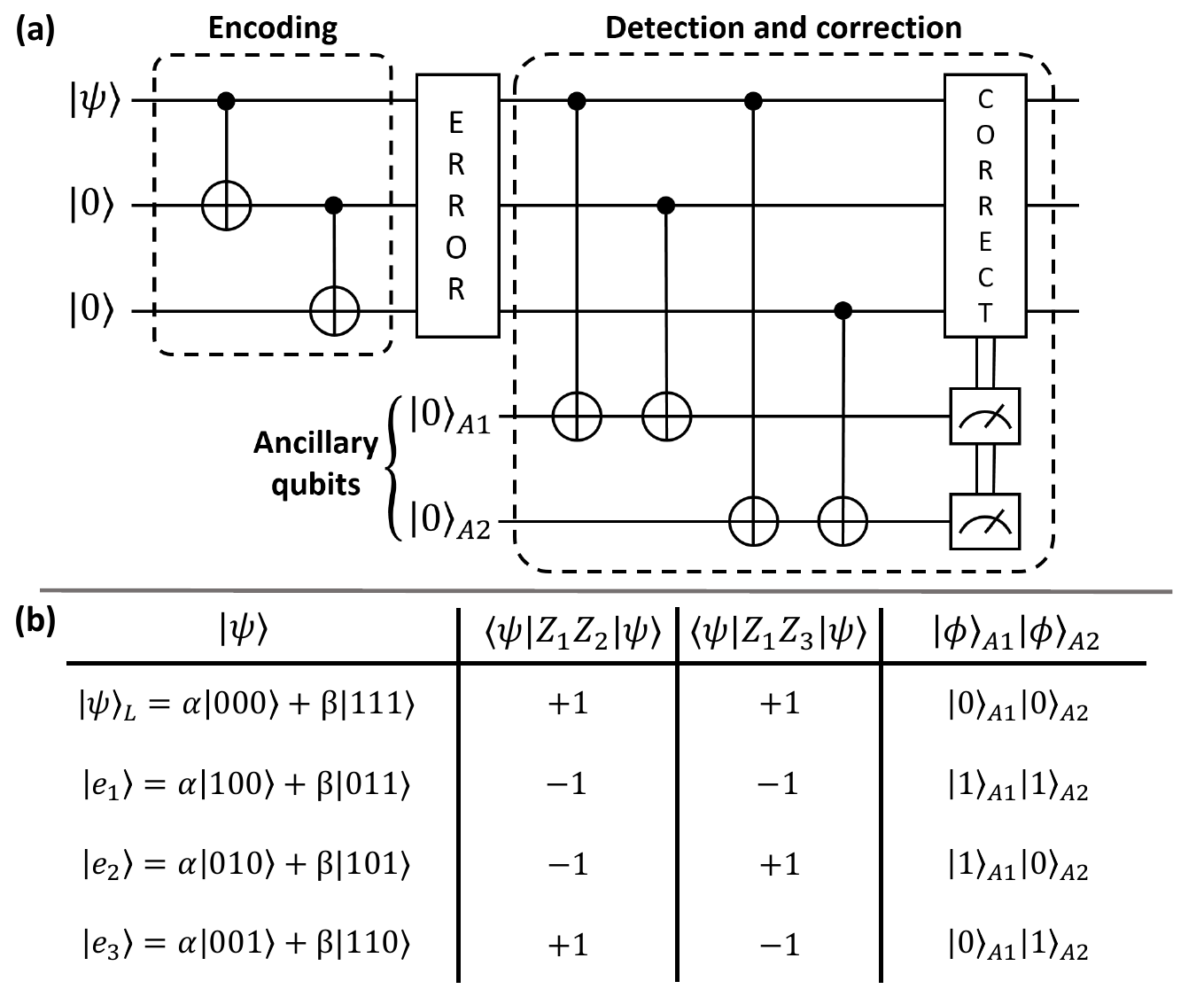}
	\caption{(a) Quantum circuit for a three-qubit code aimed at correcting a single bit flip error employing three physical qubits as the logical qubit and two ancillary qubits used for syndrome detection. Bottom table: each row represents each detectable state (no error and bit-flip on each different qubit), the other columns display the expected value on each state for both stabilizers compared to the corresponding measurement results for the ancillae.}
	\label{fig:tqc}
\end{figure}

Fortunately, the {\it redundancy} introduced by multiple qubits can still be exploited in the field of QEC. A typical toy model  to explain the basic concepts of QEC is the three-qubit code (TQC), which exploits three physical qubits to encode a single {\it logical qubit} protected from a single bit-flip error (the extension to phase errors is easy to derive, see \cite{Nielsen}). The state of the logical qubit is in fact {\it encoded} in a three-qubit state by the encoding circuit shown in the left part of Fig. \ref{fig:tqc}-(a) and consisting of two cX gates between qubits 1-2 and 2-3, where the first is the control, the second is the target.
This yields the encoded superposition of the {\it code words} $\ket{0}_L \equiv \ket{000}$ and $\ket{1}_L \equiv \ket{111}$:

\begin{equation}\label{codeword}
    \ket{\psi}_L=\alpha\ket{000}+\beta\ket{111}=\alpha\ket{0}_L+\beta\ket{1}_L.
\end{equation}

The three qubit code is then able to detect a single bit-flip error acting on this encoded qubit by performing projective measurements on two auxiliary qubits ({\it ancillae}), which are both initialized in $\ket{0}$ and are manipulated through cX gates as in Fig. \ref{fig:tqc}. The four possible results of these measurements (which by design do not affect the logical qubit state) are called {\it error syndromes} and indicate if the bit-flip error happened and on which qubit. The four different syndromes (as in the last column of Fig. \ref{fig:tqc}b) represent the four possible expectation values of the {\it stablizers} $Z_1Z_2$ and $Z_1Z_3$ (the second and the last columns). These operators have both the code words and the {\it error words} (i.e. the correctable error states $\ket{e_1}$, $\ket{e_2}$ and $\ket{e_3}$ in Fig. \ref{fig:tqc}-b) as eigenstates.
Hence,  their measurement does not destroy the logical superposition of Eq. \eqref{codeword}. As a final step the logical qubit state can be {\it corrected} by performing a X gate on the qubit affected by a bit-flip error (if detected), identified by the syndrome. This simple code allowed us to introduce the basic steps of all QEC codes: {\it encoding}, error {\it detection} via auxiliary qubits and error {\it correction}. Notably, it has been shown that multiple ions contained in a single molecular logical qubit can be exploited to implement analogous multi-qubit QEC codes \cite{ErCeEr}.

The TQC fails in case of multiple bit flip errors, i.e. the measured syndrome does not lead to the correct identification of the error. For instance, a flip of both the first two qubits would lead to the (+1,-1) syndrome [last line of the table in Fig. \ref{fig:tqc}-(b)], which would be confused with a flip of the third qubit and hence would yield the wrong correction, i.e. to a {\it logical bit flip}.
The number and type of errors that can be detected by such  multi-qubit QEC codes can be increased by using a larger number of physical qubits encoding a single logical qubit and often many {\it ancillae} \cite{Terhal2015,Devitt2013,Mariantoni2012,Google_surface}. However, such an approach quickly becomes very demanding on the hardware, because it involves an impressive number of qubits and of two-qubit gates (often between distant physical qubits) to actually implement correction and quantum operations\cite{PRX_multiqubitscaling}. 

A possible alternative to multi-qubit  codes is given by the {\it qudit} approach, in which an error-protected logical qubit is encoded in a single multi-level quantum system, intrinsically providing the additional degrees of freedom required by QEC. An interesting example of such systems are molecular spin qudits, whose unparalleled degree of control at the synthetic level can be exploited to design error-resistant units (see Sec. \ref{sec:decoqudit}) and also to embed QEC. To achieve this, we need an efficient {\it encoding} of the logical qubit in a multi-level structure, ancillary systems that can be measured to {\it detect} the error without collapsing the logical superposition, and the possibility of {\it correcting} the error on the encoded state. 

\begin{figure*}[ht]
    \centering
	\includegraphics[width=0.95\textwidth]{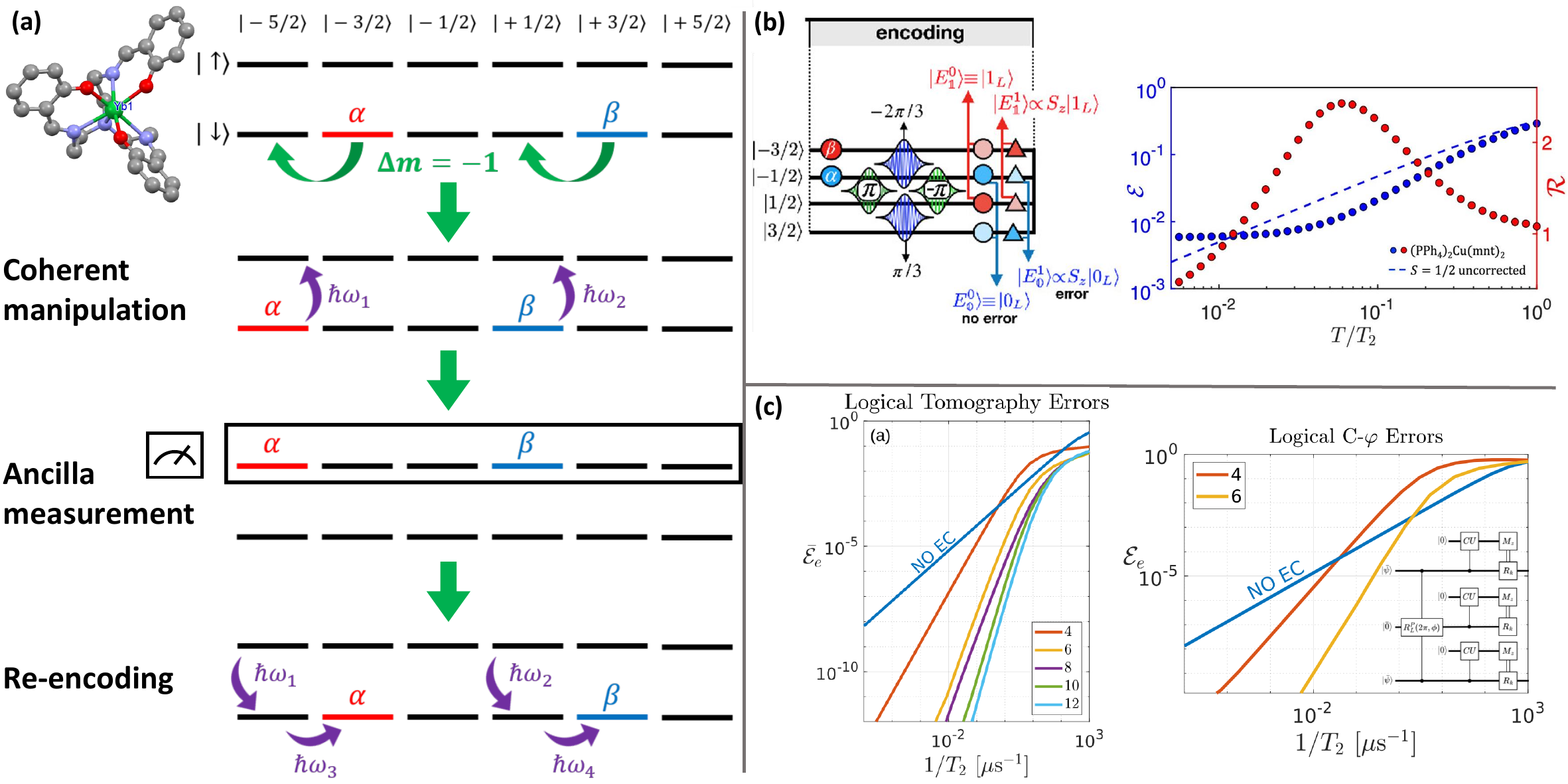}
	\caption{(a) Qudit code protecting against amplitude-shift errors, proposed \cite{jacsYb} for a nuclear $I=5/2$ spin qudit coupled to an effective electronic spin doublet, such as $^{173}$Yb(trensal) (inset). The steps for detecting a $\Delta m_I = -1$ error are sketched: coherent excitation of the ancilla for the two transitions corresponding to an error actually occurred on the qudit; measurement of the electronic ancilla without collapsing the nuclear spin superposition; correction if the ancilla is found in its $\ket{\uparrow}$ state. (b) Left: sequence of pulses to encode a logical qubit protected to first order from pure dephasing into a spin $S=3/2$. Full control of the qudit can be achieved by pulses resonant with $\Delta m =\pm 1$ transitions. Right: error after a memory time $T/T_2$ followed by error-correction (dots), compared with an uncorrected spin 1/2 (dashed line) and ratio between the two (red axis and dots), representing the gain of the correction procedure. The simulation, is performed on a nuclear spin qudit coupled to an electronic spin doublet \cite{BaderNatComm14}. The correction procedure is not fault-tolerant, and hence for small values of $T$ the simple spin 1/2 outperforms the encoded qubit. Both figures are reprinted with permission from \cite{JPCLqec}. (c) Fault-tolerant procedure to implement logical gates followed by error correction against pure dephasing. Left: error after single qubit gates, averaged over various planar rotations. Right: error after a two-qubit logical controlled-phase gate. For values of the elementary dephasing rate $1/T_2$ below a certain threshold (intersection between blue and yellow/orange curves), the corrected logical qubit outperforms the spin 1/2, i.e. the final error is smaller. The correcting power increases by using an increasing number of levels (shown in the legends) for the encoding, as evidenced by the higher slope of the curves.  Both figures are reprinted with permission from \cite{Mezzadri}.  }
	\label{fig:qudit_QEC}
\end{figure*}

To illustrate the basic principles of the qudit encoding, we consider a simple code protecting against an amplitude shift error, which represents an extension of the bit flip error considered above for qubits.
The qudit embedding an error-protected qubit is a 6-level system, coupled to a 2-level ancilla. This idea was first proposed \cite{jacsYb} for $^{173}$Yb(trensal) $I=5/2$ nuclear qudit, coupled to its $s=1/2$ electronic spin via hyperfine interaction [see Eq. \eqref{eq:nuclei}]. The logical qubit state $\ket{\psi}=\alpha\ket{0}+\beta\ket{1}$ is encoded as follows [see Fig. \ref{fig:qudit_QEC}-(a)]:
\begin{equation}\label{eq:encodedqudit}
\ket{\psi}_L=\alpha\ket{-3/2}\otimes\ket{\downarrow}+\beta\ket{3/2}\otimes\ket{\downarrow}=\alpha\ket{0}_L+\beta\ket{1}_L,
\end{equation}
where the system eigenstates are practically factorized in the nuclear  $\ket{m_I=-5/2, ..., 5/2}$ and electronic $\ket{m_s = \downarrow, \uparrow}$ spin projection along the external field. The logical qubit is encoded in the subspace in which the electronic ancilla is $\ket{\downarrow}$ and exploiting only two of the 6 levels of the qudit.
Amplitude shift errors induce a $\Delta m_I=\pm1$ shift on the qudit state, producing one of the following error words:
\begin{align}
\ket{e_-} &= \alpha\ket{-5/2}\otimes\ket{\downarrow} + \beta\ket{1/2}\otimes\ket{\downarrow}, \\
\ket{e_+} &= \alpha\ket{-1/2}\otimes\ket{\downarrow} + \beta\ket{5/2}\otimes\ket{\downarrow}.
\end{align}
Hence, depending on the possible error that occurred, the system can now be found in one of the three mutually orthogonal states $\ket{\psi}_L$, $\ket{e_-}$ or $\ket{e_+}$.
{\it Detection} is made possible by measuring the state of the 2-level ancilla after its qudit-state dependent excitation. Indeed, transitions of the ancilla can be selectively addressed based on the qudit state, thanks to the hyperfine interaction. In particular, by  driving both the  $\ket{-5/2}\otimes\ket{\downarrow}\leftrightarrow\ket{-5/2}\otimes\ket{\uparrow}$ and $\ket{1/2}\otimes\ket{\downarrow}\leftrightarrow\ket{1/2}\otimes\ket{\uparrow}$ transitions via two independent resonant pulses, we obtain an excitation of the ancilla only if a $\Delta m_I=-1$ amplitude shift error has taken place, while still maintaining the correct superposition. Then, performing a projective measurement only on the electronic ancilla can give two different results: 
\begin{enumerate}
    \item $\ket{\uparrow}$: a $\Delta m_I=-1$ amplitude shift error is detected, and with a sequence of tailored pulses it is possible to retrieve the correct encoded state \eqref{eq:encodedqudit};
    \item $\ket{\downarrow}$: either a $\Delta m_I=+1$ amplitude shift error has taken place or the qudit is still in the correct encoded state. At this point, repeating the same detection procedure for the $\Delta m=+1$ amplitude shift error (two pulses addressing the $\ket{-1/2}\otimes\ket{\downarrow}\leftrightarrow\ket{-1/2}\otimes\ket{\uparrow}$ and $\ket{5/2}\otimes\ket{\downarrow}\leftrightarrow\ket{5/2}\otimes\ket{\uparrow}$ transitions followed by a projective measurement of the ancilla) allows one to finally determine which error occurred and then correct the state, if needed.
\end{enumerate}
This detection procedure is summarised in Fig. \ref{fig:qudit_QEC}-(a). Analogously to multi-qubit codes, measurements of the ancilla give the three different {\it error syndromes} of qudit amplitude shift errors: first measurement $\ket{\uparrow}$, first measurement $\ket{\downarrow}$ then second measurement $\ket{\uparrow}$, and both measurements $\ket{\downarrow}$.

This simple example highlights the fundamental steps to design a QEC code: (i) focus on the errors we aim to correct. This point is crucial, since any QEC code can only correct a subset of the errors occurring on the hardware. Hence, it is mandatory to identify a hierarchy between errors and correct the most important ones. Then we need to find code words satisfying essentially two conditions: "correctable" errors must (ii) preserve the encoded superposition (i.e. do not alter $\alpha$ and $\beta$) and (iii) do not induce an overlap between the code words $\ket{0}_L$ and $\ket{1}_L$.
These requirements are formalised by the {\it Knill-Laflamme} conditions \cite{KnillLaflamme}:
\begin{align}\label{eq:KnillLaflamme}
   \braket{0_L\vert E_kE_j^\dagger\vert0_L} &= \braket{1_L\vert E_kE_j^\dagger\vert1_L},\\
   \braket{0_L\vert E_kE_j^\dagger\vert1_L} &= 0,
\end{align}
where $E_{k,j}$ are any pair of error operators corrected by the code. The first equation ensures that under the effect of any correctable error the quantum superposition is not disrupted, while the second that $\ket{0_L}$ and $\ket{1_L}$ remain orthogonal. \\
In the case of the amplitude-shift code presented above, we can easily check that $\ket{0}_L \equiv \ket{-3/2}\otimes \ket{\downarrow}$ and $\ket{1}_L \equiv \ket{3/2}\otimes \ket{\downarrow}$ satisfy the Knill-Laflamme conditions for error operators $E_{\pm} = \sum_{m_I} \ket{m_I\pm1} \bra{m_I}$. 
The same ideas can be exploited to protect against other errors, more relevant for MNMs.
In particular, as discussed in Section \ref{sec:deco}, dephasing is the main source of decoherence in molecular spins. Hence, QEC codes aimed at correcting phase errors are the most valuable in quantum computing with molecular spins. In particular, a 4-level qudit (such as a spin $3/2$) can be exploited to encode a qubit protected from the first order effect of pure dephasing. This translates into identifying code words fulfilling Knill-Laflamme conditions for the two error operators $\{I, S_z\}$.
A possible choice is given by the following superpositions of $\ket{m}$ states \cite{JPCLqec}:
\begin{align}\label{eq:encoding3/2}
    \ket{0_L} &= \frac{\sqrt{3}}{2}\ket{-1/2} + \frac{1}{2}\ket{3/2} \\
    \ket{1_L} &= \frac{1}{2}\ket{-3/2} + \frac{\sqrt{3}}{2}\ket{1/2}.
\end{align}

Starting from a qubit state defined as a generic two level superposition (e.g $\ket{\psi_0}=\alpha\ket{-3/2}+\beta\ket{-1/2}$) the encoded state $\ket{\psi_L}=\alpha\ket{0_L}+\beta\ket{1_L}$ can be prepared with a sequence of $\Delta m = \pm 1$ pulses \cite{JPCLqec,ChizziniPCCP}, as shown in panel (b) of Figure \ref{fig:qudit_QEC}. Indeed, the capability of driving  $\Delta m = \pm 1$ transitions ensures universal control of the qudit state \cite{Chizzini2022,Planar_rot}. An $S_z$ error (occurring with a certain probability) will then yield the transformation
\begin{align}\label{eq:encoding3/2}
    \ket{0_L} \rightarrow \ket{e_0} &= -\frac{1}{2}\ket{-1/2} + \frac{\sqrt{3}}{2}\ket{3/2} \\
    \ket{1_L} \rightarrow \ket{e_1} &= -\frac{\sqrt{3}}{2} \ket{-3/2} + \frac{1}{2}\ket{1/2},
\end{align}
thus acting on the encoded superposition as follows
\begin{equation}
    \ket{\psi_L} \rightarrow \ket{\psi_1} = \alpha \ket{e_0} + \beta \ket{e_1}.
\end{equation}
We note that $\ket{\psi_1}$ preserves the original coefficients $\alpha$ and $\beta$ and is orthogonal to $\ket{\psi_L}$, i.e. $\{ \ket{0_L}, \ket{1_L}, \ket{e_0}, \ket{e_1}\}$ form an orthonormal basis set of the qudit $S=3/2$ Hilbert space. Hence, by projecting either onto the {\it logical} $\{ \ket{0_L}, \ket{1_L} \}$ or onto the {\it error subspace} $\{ \ket{e_0}, \ket{e_1}\}$ we can detect the possible occurrence of an $S_z$ error without collapsing the logical superposition. These projections are formally represented by operators
\begin{eqnarray}\nonumber
    P_L &=& \ketbra{0_L}{0_L} + \ketbra{1_L}{1_L} \\
    P_e &=& \ketbra{e_0}{e_0} + \ketbra{e_1}{e_1} .
    \label{eq:projections}
\end{eqnarray}
The simplest way to implement them is by first using proper sequence of $\Delta m = \pm 1$ pulses to map back each superposition in the set $\{ \ket{0_L}, \ket{1_L}, \ket{e_0}, \ket{e_1}\}$ into a $\ket{m}$ eigenstate of the qudit. For instance, $\ket{0_L}\rightarrow \ket{3/2}$, $\ket{e_0}\rightarrow \ket{1/2}$, $\ket{1_L}\rightarrow \ket{-1/2}$, $\ket{e_1}\rightarrow \ket{-3/2}$. Then, a spin 1/2 ancilla interacting with the qudit can be exploited for error detection similarly to the amplitude shift code described above. In particular, excitations of the ancilla are made dependent on the state of the qudit thanks to the mutual coupling and can be induced by resonant pulses. To preserve the logical superposition ($\alpha, \beta$), two pulses are sent simultaneously (e.g. for the qudit being in $\ket{3/2}$ and $\ket{1/2}$) and finally the state of the ancilla is measured independently from that of the qudit. \\
A similar encoding can be extended to higher half-integer spins $S$, thus correcting higher powers of $S_z$ \cite{JPCLqec}. In practice, at most an additional error can be corrected for each added pair of levels in the qudit. This improves the correcting power of the code.

Panel (c) of Figure \ref{fig:qudit_QEC} (taken from [\onlinecite{JPCLqec}]) shows the simulated performance of this procedure by considering a nuclear qudit spin $I=3/2$ coupled to its electronic $S=1/2$ ancilla via hyperfine coupling, as in the Cu$^{2+}$ ion of the molecular qubit characterized in \cite{BaderNatComm14}. In this plot the blue dots represent the error after the implementation of the code, defined as $\mathcal{E}=1-\bra{\psi_L}\rho\ket{\psi_L}$, where $\rho$ is the density matrix obtained at the end of the simulation and $\ket{\psi_L}$ is the correct encoded state. This quantity is compared to the time evolution of an uncorrected spin $1/2$ subject to pure dephasing (dashed line) as a function of the idle time between encoding and detection (memory time). The poor performance of QEC at short memory times, represented by the initial error plateau, arises from dephasing errors occurring during the sequence of pulses used to detect and correct. These additional errors are not corrected by the code and overcome correctable ones at short memory times. 
This trend is inverted for longer memory times, where the advantage of QEC becomes evident.

The extra errors introduced by the correction procedure do not fulfil Knill-Laflamme conditions and hence are not managed by the code. This problem can be circumvented by designing {\it error-transparent} pulse sequences \cite{Reinhold2020,Puri2020,Error_transparent}, which do not affect the correctable set of errors. It was recently found that a way to achieve this is to exploit the all-to-all connectivity between qudit eigenstates, i.e. the capability to induce transitions between all states without the limitation of $\Delta m=\pm 1$. This connectivity can be obtained by properly chosen MNMs, for instance by  considering multi-spin systems with competing anti-ferromagnetic interactions, where the several low-energy total spin multiplets can be connected by natural presence of anisotropic terms or symmetry distorsions in the spin Hamiltonian \cite{Mezzadri}. As explained in Section \ref{sec:decoqudit}, this type of systems is also associated to favourable decoherence patters, which would further improve the efficiency of qudit-based QEC. Figure \ref{fig:qudit_QEC}-(d) (taken from \cite{Mezzadri}) shows that the all-to-all-connectivity of these systems enables not only transparent error correction, but also  a universal set of {\it error transparent} gates between logical qubits, thus potentially leading to  {\it fault-tolerant} quantum computation.



\section{Readout}
\label{sec:readout}
\begin{figure*}[ht]
    \centering
	\includegraphics[width=0.98\textwidth]{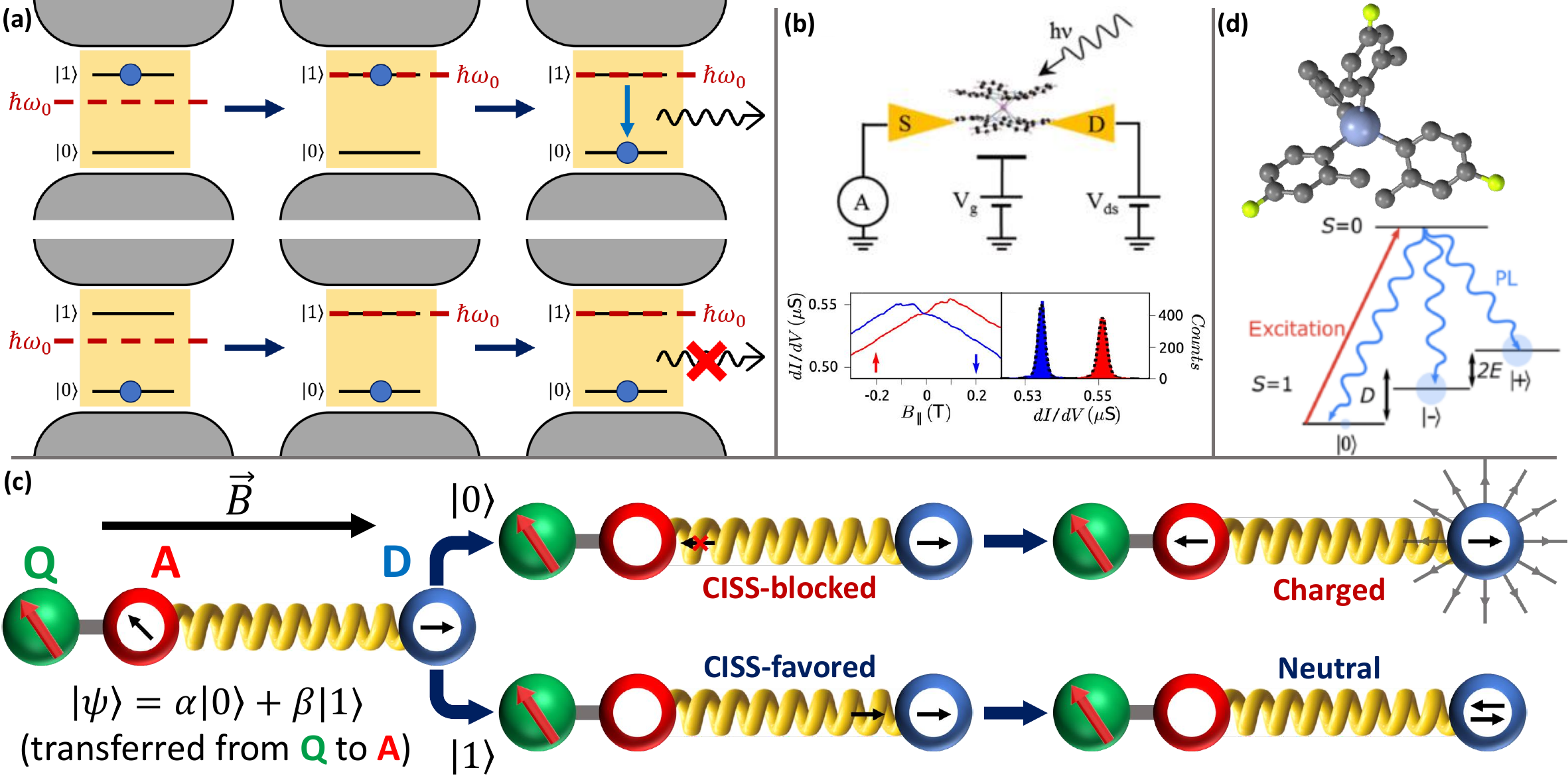}
	\caption{Different proposed methods for the readout of individual molecular spin qudits. (a) Coupling the spin to an empty superconducting resonator and then swapping the spin state into the photon field by tuning the photon frequency in resonance with the spin gap. Top: if the qubit is in the excited state, a photon is emitted and detected by a single-photon counter \cite{Chiesa2023prappl,Walter2017}. Bottom: if the qubit is in the ground state, no photon is emitted (energy is practically conserved) nor detected. (b) Exploiting a single-molecule transistor bound to a conducting part of TbPc$_2$ molecule to assess the spin state of Tb ion \cite{Godfrin2017b}. Bottom: differential conductance as a function of the magnetic field and related histogram of differential conductance values. The exchange interaction between the qubit spin and the read-out quantum dot yields shifts the  conductance signals depending on the two spin orientations (red/blue).  Reprinted with permission from C. Godfrin et al., Phys. Rev. Lett. {\bf 119}, 187702 (2017) \cite{Godfrin2017}, Copyright (2017) by the American Physical Society and from C. Godfrin et al., ACS Nano {\bf 11}, 3984 (2017) \cite{Godfrin2017b}, Copyright(2017) by the American Chemical Society. (c) Use optical transitions of molecular colour centres. Molecular structure of the chromium-based molecular colour center Cr-(IV)($\sigma$-tolyl)$_4$ and level diagram with optically pumped transition (red) and decay to the $m=\pm 1$ excited states.  Reprinted with permission from S. L. Bayliss et al., Phys. Rev. X {\bf 12}, 031028 (2022) \cite{Bayliss2022}, Copyright (2022) by the American Physical Society (d) Conversion of spin information into charge information via the CISS effect and subsequent readout of the charge state of a radical spin pair. The molecular qubit state $\ket{\psi}$ is initially transferred to the acceptor via resonant pulses \cite{Chiesa2023,Chiesa2021,JCP_CISS}, then spin-selective charge recombination through the chiral bridge leads to different charge states based on the probability of the acceptor spin of being in $\ket{0}$ or in $\ket{1}$.  } 
	\label{fig:readout}
\end{figure*}
At the end of a quantum algorithm (and in some cases also during it) one needs to access the state of the register by measuring the qubits in a given basis. This measurement is usually projective in the quantum mechanical sense. This means that if we consider the observable $A = \sum_k a_k \ket{k} \bra{k}$ (with $\ket{k}$ the eigenvectors of $A$ and $a_k$ its discrete eigenvalues) and we get $a_k$ as output, the system state-vector is projected onto
\begin{equation}
	\ket{\psi} \rightarrow \frac{P_k \ket{\psi}}{\norm{P_k\ket{\psi}} } 
\end{equation}
with probability $|\braket{k|\psi}|^2$.
Measuring the state of a qubits register corresponds, in most architectures, to measuring the local observable $\sigma_z$ on each qubit. This yields $\ket{0}$ and $\ket{1}$ as possible outcomes with probabilities $|\braket{0|\psi}|^2 = |\alpha|^2$ and $|\braket{1|\psi}|^2 = |\beta|^2$. The corresponding value of $\langle \sigma_z \rangle$ is given by the difference of these two probabilities, resulting in $|\alpha|^2-|\beta|^2$.
The off-diagonal elements of the qubit density matrix can be accessed by performing a $R_y(\pi/2)$ or $R_x(\pi/2)$ rotation on the qubit before the measurement in the z-basis, thus obtaining the real and imaginary part of the coherence, respectively. This is equivalent to measuring $\langle \sigma_x \rangle$ and $\langle \sigma_y \rangle$. For instance, a $R_y(\pi/2)$ transforms $\ket{\psi}=\alpha \ket{0} + \beta \ket{1}$ into
\begin{equation}
\ket{\psi_x} = \frac{\alpha+\beta}{\sqrt{2}} \ket{0} - \frac{\alpha-\beta}{\sqrt{2}} \ket{1} . 
    \label{eq:xmeas}
\end{equation}
A $z-$basis measurement gives $\ket{0}$ or $\ket{1}$ with probabilities $|\alpha\pm\beta|^2/2$, whose difference is exactly $\langle \sigma_x \rangle$ for the original state, i.e. $\alpha \beta^* + \alpha^* \beta = 2 {\rm Re} [\alpha \beta^*]$. Analogously one can obtain $\langle \sigma_y \rangle$ by applying an $R_x(\pi/2)$ before measuring. \\

Projective readout on a molecular spin quantum register requires to access the state of a single spin (i.e. a tiny magnetic moment), which represents one of the most important challenges in the field. Different lines are currently being pursued to overcome this hurdle: (i) strongly coupling a single molecular spin to the photon in a superconducting resonator. (ii) Exploiting a spin-to-charge conversion mechanism to get a much larger and readable signal, either by means of conducting properties of molecules bound to a single-molecule transistor \cite{Thiele2014} or via the recently proposed chirality-induced spin selectivity effect \cite{Chiesa2023}. (iii) Use molecular color-centers with proper spin-selective optical transitions \cite{Bayliss2022}.
The ideas behind these different possibilities are sketched in Fig. \ref{fig:readout}. 

Scheme (i) follows the approach developed for the control and readout of superconducting qubits, already outlined in Sec. \ref{sec:2qubits}.B. Compared to that platform, the bottleneck here is represented by reaching the strong coupling between a single molecular spin and the photon field within the resonator. Once achieved, this would also allow for a projective measurement of a single spin. The most efficient scheme is outlined in Fig. \ref{fig:readout}-(a): start with an empty resonator (no photons) and then tune its frequency to match the gap between the two states of the spin qubit. Then a photon is emitted only if the qubit was in the excited state. Finally, a single-photon sensor detects the possible presence of the photon, thus reading out the former state of the qubit. This resonant scheme would enable a fast readout of the state of a single spin qubit even in presence of a not very large spin-photon coupling \cite{Chiesa2023prappl}. It can also be generalized to a spin qudit, by selectively addressing one gap at a time.

Scheme (ii) aims to magnify the tiny spin information by converting it into a charge information. Note that even the charge of a single electron generates a rather strong electric field nearby, which can be detected by different types of electro-meters.  
Along this line, a first method relies on conducting properties of phtalocyanine ligands in the TbPc$_2$ molecule shown in Fig. \ref{fig:readout}-(b) which are exchange-coupled to the effective electronic spin 1/2 of the Tb ion. Hence, if the molecule is placed between two-electrodes, the  conductance through it shows steps each time the electronic spin reverses \cite{Thiele2014}.  \\
Another approach was recently proposed, based on the chirality-induced spin selectivity (CISS) effect in electron-transfer reactions \cite{Eckvahl2023}. Due to the CISS effect, electrons transferred through a chiral bridge are spin-polarized. The reaction can be controlled by external stimuli, usually by photo-exciting the electron donor. Then, only electrons characterized by a given projection of the spin along the chiral axis are transferred. The charge transfer yields a significant change in the electro-static potential, which can be detected e.g. by a nearby quantum dot \cite{Yang2011} similarly to the setup of Ref. \cite{Thiele2014}. Conversely, 
electrons with the opposite spin projcetion are not transferred, resulting in a very different electro-static potential on the dot.
Hence, the process acts on the spin degree of freedom as a projective readout. 
A possible setup exploiting the CISS effect for quantum computation was proposed in \cite{Chiesa2023} and is reported in Fig. \ref{fig:readout}-(c) with a molecular spin qubit (Q) coupled to a donor-chiral bridge-acceptor unit (D$-\chi-$A in Fig. \ref{fig:readout}-(c)). In that case, a charge separated state with well-defined spin polarization was generated by photo-exciting the donor. After electron transfer, the spin polarization can be swapped from the acceptor to the qubit linked to it via a simple sequence of resonant microwave pulses, thus initialising it even at high temperature. 
At the end of the computation, the generic state of the qubit $\ket{\psi}=\alpha \ket{0} +\beta \ket{1}$ is first swapped back to the acceptor. Then, readout is  achieved by inducing spin-selective recombination (thanks to the CISS effect) by an electric field or light. As shown in Fig. \ref{fig:readout}-(c),  only one of the two spin projections is allowed to recombine by the CISS effect (bottom), while the other is blocked (top). This leads to very different electrostatic potentials in the vicinity of the radical pair (a neutral system vs an electric dipole), which can be more easily discriminated.

Scheme (iii) is based on $S=1$ systems (such as Cr$^{4+}$ compounds) with an $m=0$ ground state and a $m=\pm 1$ excited doublet with small $D$ and $E$, similarly to NV centers, as shown in Fig. \ref{fig:readout}-(d) \cite{Freedman_Science,Bayliss2022}. Conversely, due to the strong ligand-field the lowest-energy excited electronic state is a singlet $S=0$. This yields narrow optical transitions between the ground $S=1$ and the excited $S=0$ multiplet. Initialisation and readout are implemented as follows. 
If the transition between the ground state and the excited $S=0$ state is optically pumped, the state decays only to the $m=\pm 1$  due to optical selection rules in spontaneous emission, thus in fact initialising the qubit. 
In a symmetric way, the qubit state can also be read-out, because the probed $m$ state within the $S=1$ ground multiplet (e.g. $m=0$) will give rise to higher photo-luminescence [PL in Fig. \ref{fig:readout}-(d)]. Although implemented so far only on crystals \cite{Freedman_Science,Bayliss2022,Serrano2022}, it should be possibile to extend the use of this technique to the single molecule level.

We finally note that most of the proposed methods for readout can also be used to initialize the spin state of the qubit. More generally, a projective measurement of the state of the qubit is also a way of initializing its state. 
This method can be faster than thermal initialization to the ground state and can often be applied at higher temperatures.  \\

\section{Examples of quantum algorithms}
\label{sec:algorithms}
As illustrated in the previous sections, 
MNMs can be employed as building blocks of a general purpose quantum computer, i.e. a device which allows for the implementation of any quantum algorithm. 
However, important challenges must still be overcome before approaching this goal and devices available in the near future will still be noisy and of limited size. In this perspective, it is important to design hardware-efficient strategies to mitigate errors and improve the capabilities of current platforms. \\
As mentioned in Sec. \ref{sec:deco}.B, the intrinsic multi-level structure of MNMs provides an important tool to go beyond the binary logic and design algorithms which explicitly take advantage from it.

\subsection{Grover's algorithm}
This approach was pursued for the first time in the early 2000s \cite{Loss2001,Loss2003}, when two schemes where proposed to implement the Grover's search algorithm on single MNM. Grover's algorithm exploits superposition and interference of quantum states to perform a search into an unstructured database more efficiently than a classical device. In both schemes, the idea was to encode the database entries into the $2S+1$ levels of a single total spin $S$ molecule, such as the single-molecule magnets Fe$_8$ or Mn$_{12}$ or a nuclear spin $I\ge 1$. Then, a superposition of all of these states was generated and finally only the searched state was amplified by  sequences of electromagnetic pulses of proper amplitude, duration, frequency and phase. \\
\begin{figure}[t]
    \centering
    \includegraphics[width=0.47\textwidth]{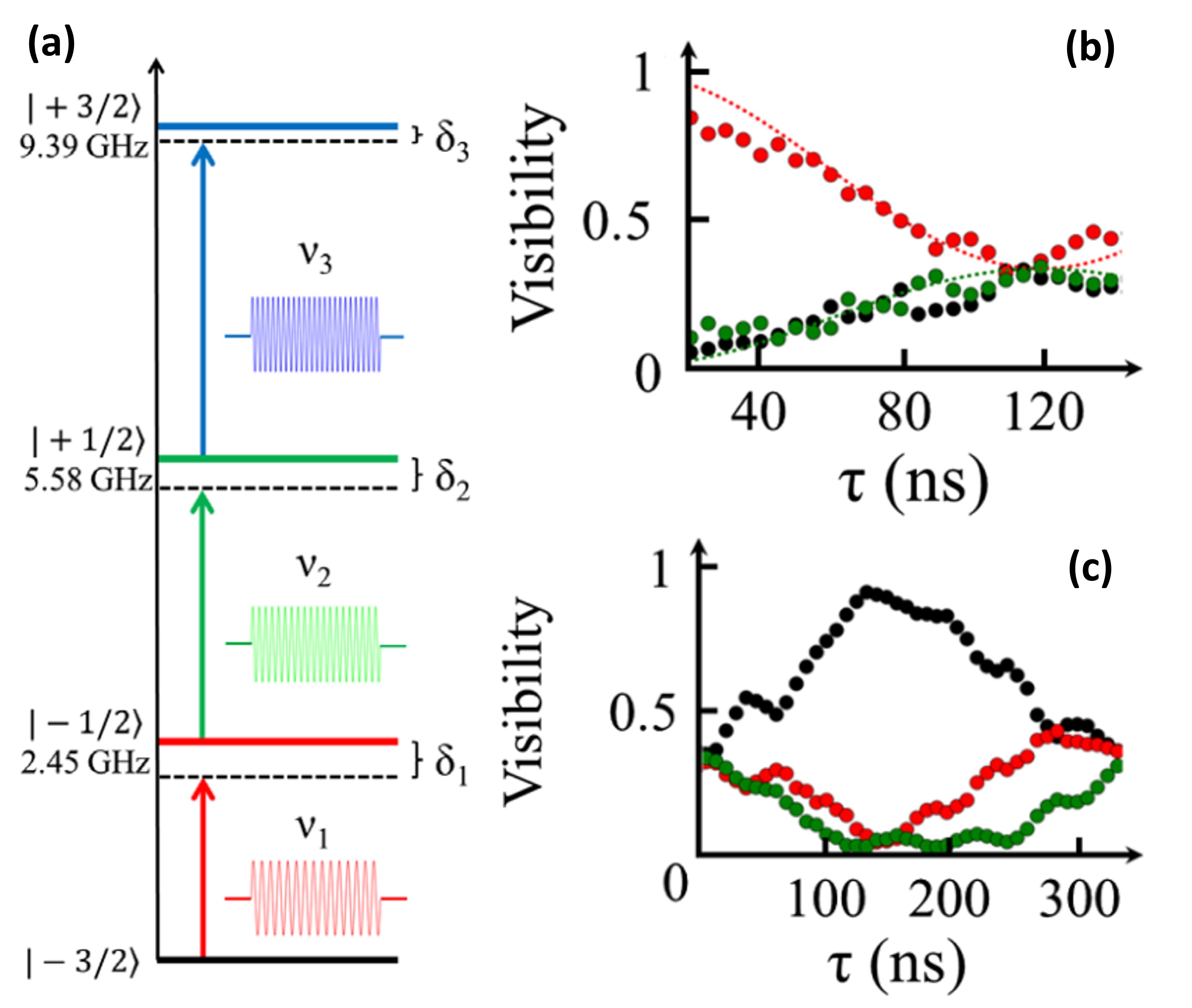}
    \caption{(a) Energy level diagram of the $I=3/2$ nuclear spin qudit, with transitions between different $m_I$ states well resolved by the large hyperfine and quadrupolar interactions. Full control of the system is achieved with $\Delta m_I =\pm 1$ pulses of frequency $\nu_i$ and detuned by an amount $\delta_i$ from the respective spin energy gap. By tuning $\nu_i$, $\delta_i$ and the phases of the pulses and let the system evolve under the effct of this multi-frequency pulse for the suitable amount of time $\tau$, first a superposition of three nuclear spin states (different colours) with equal amplitudes is generated (b) and then the searched state (here the black one) is amplified (c). 
    Reprinted with permission from C. Godfrin et al., Phys. Rev. Lett. {\bf 119}, 187702 (2017) \cite{Godfrin2017}, Copyright (2017) by the American Physical Society.}
    \label{fig:Grover}
\end{figure}
The idea was applied later on to a nuclear spin qudit into a TbPc$_2$ molecule \cite{Godfrin2017}, in the single-molecule transistor setup illustrated above. Results are shown in Fig. \ref{fig:Grover}. The Grover algorithm is implemented by applying a three-frequency pulse to the system, where the frequencies and phases are chosen to obtain the desired dynamics after a suitable time interval $\tau$. In particular, first an equal-weight superposition of three nuclear states is created [Fig. \ref{fig:Grover}-(b)] and then the amplitude corresponding to the searched state is amplified  [Fig. \ref{fig:Grover}-(c)].

\subsection{Digital Quantum Simulation} 

Another class of problems which attracted a considerable attention is represented by  Quantum Simulation (QS), with several proposals to implement it on existing molecular units \cite{Santini2011,SciRepNi,Chem,VO2}. 
Quantum simulation is generally recognized as one of the closest steps to demonstrate a quantum advantage over classical hardwares. Indeed, simulating the dynamics of a quantum system on a register of a few dozens of qubits (even without quantum error correction) would already represent a huge achievement.

Furthermore, in most cases at the end of the QS we are interested in extracting expectation values of target observables. Hence, QS experiments can be realised on a quantum hardware consisting of an ensemble of MNMs, i.e. a collection of practically identical molecules ordered in a crystal lattice. In this respect, we note that the most important contribution to decoherence of the ensemble is given by inhomogeneous broadening. Indeed, different molecules in the ensemble will be characterised by slightly different spin Hamiltonian parameters and hence slightly different energy gaps. For an ensemble of spins 1/2 preceding under the effect of an external field, this translates into a slightly different precession frequency for each spin. As a consequence, even if the ensemble is initialised in a state with all the spins parallel, the initially large total magnetisation of the ensemble quickly averages to 0. The standard approach to tackle this problem is the Hahn-echo pulse sequence \cite{Hahn}. To detect the spin state of the system, one first applies a $\pi/2$ pulse to project the longitudinal component of the magnetisation (which we aim to detect) into the plane perpendicular to the field. Then the system is let free to precede for a time interval $\tau$, during which different spins dephase because of inhomogeneous broadening. Then, a $\pi$ pulse is applied to {\it re-focus} the spin state, i.e. the state of each spin is reversed and consequently (if the Hamiltonian is time independent) also its time evolution. Therefore, at time $2 \tau$ an echo is detected. This idea can be applied also to more complex spin systems and allows one to compensate {\it static} inhomogenities in the spin Hamiltonian parameters. As discussed in Sec. \ref{sec:deco}, this method cannot correct errors induced by a fluctuating spin bath. 

Implementing the quantum simulation of a target Hamiltonian $\mathcal{H}$ involves the following steps: (i) map the degrees of freedom of the target problem into the hardware register; (ii) re-write the target Hamiltonian into the hardware, mapping $\mathcal{H}\rightarrow H$;
(iii) decompose the resulting time-evolution operator $U(t) = \exp [-i H t]$ into a sequence of gates which can be implemented on the hardware. 
All these steps can be implemented either exactly or with controllable approximations. In general, one needs to find a trade-off between the "digital" error due to such approximations and the number of gates or qubits, which cannot be increased beyond the capabilities of the hardware. \\
For the quantum simulation of spin 1/2 Hamiltonians, steps (i) and (ii) are straightforward and exact. Conversely, step (iii) usually cannot be implemented directly, because the target multi-qubit Hamiltonian contains different terms (e.g. $H=H_{1}+H_{2}$) which in general do not commute. This problem is faced by relying on the Suzuki-Trotter approximation:
\begin{equation}
	U(t) = e^{-i(H_1+H_2)t} \approx \left( e^{-i H_1 t/n} e^{-i H_2 t/n} \right)^n ,
	\label{eq:Suzuki}
\end{equation}
where the error is of order $(t/n)^2$ and can be reduced arbitrarily by increasing $n$, i.e. by applying $n$ times evolution operators $U(t/n)$ acting for sufficiently small time-slices.
Then, each term in the product is recast into elementary one- and two-body gates implementable on the hardware.

Let us consider a simple example to illustrate this point, namely the QS of the  transverse-field Ising model on a pair of spins. The target Hamiltonian is 
$\mathcal{H} = b (s_{x1}+s_{x2}) +  J s_{z1} s_{z2}$, where one can identify $H_1 = b (s_{x1}+s_{x2})$ and $H_2 = J s_{z1} s_{z2}$, with $[H_1, H_2] \neq 0$.
We can easily find a decomposition of both unitary operators $U_i(\tau) = \exp[-i H_i \tau]$ in one- and two-qubit gates as follows:
\begin{equation}
	U_1(\tau) = e^{-i (s_{x1}+s_{x2}) b\tau} = R_x^{(1)} (b\tau) R_x^{(2)} (b\tau)
\end{equation}
and 
\begin{equation}
	U_2(\tau) = e^{-i s_{z1} s_{z2} J \tau} = U_{cJ\tau} R_z^{(1)} (J\tau/2) R_z^{(2)} (J\tau/2),
 \label{eq:zz}
\end{equation}
where we have indicated by $R_\alpha^{(i)} (\theta)$ a rotation about axis $\alpha$ of an angle $\theta$ on qubit $i$ and by $U_{c\varphi}$ a controlled-phase gate with a phase $\varphi$.
The implementation of single qubit gates about arbitrary axes and of controlled-phase gates on a molecular spin quantum processor have already been illustrated in Sec. \ref{sec:qubits} and \ref{sec:2qubits}. Here we note that the $U_2(\tau)$ operator could also be implemented directly [without needing decomposition \eqref{eq:zz}], in the same setup described in Sec. \ref{sec:2qubits}, by exploiting two semi-resonant excitations of the switch for both the two states of the logical qubits pair $\ket{00}$ and $\ket{11}$. Indeed, this would add a relative phase between parallel and anti-parallel configurations of the qubit pair, exactly as the Ising evolution operator $U_2(\tau)$. \\
Eq. \eqref{eq:zz} can be easily generalised to simulate the time evolution induced by a generic spin-spin interaction of the form $s_{\alpha1} s_{\beta 2}$, by exploiting simple algebraic properties of Pauli matrices. For instance
\begin{eqnarray}
R_x \left(-\frac{\pi}{2}\right) \sigma_z R_x \left(\frac{\pi}{2}\right) &=& \sigma_y \\
R_y \left(\frac{\pi}{2}\right) \sigma_z R_y \left(-\frac{\pi}{2}\right) &=& \sigma_x ,
    \label{eq:paulieq}
\end{eqnarray}
from which it is easy to show that
\begin{equation}
e^{-i s_{x1} s_{x2} \varphi} =  
R_y \left(\frac{\pi}{2}\right) e^{-i s_{z1} s_{z2} \varphi} R_y \left(-\frac{\pi}{2}\right) 
\end{equation}
with the short-hand notation $R_y(\theta) = R_y^{(1)} (\theta) R_y^{(2)} (\theta)$, and similarly for other spin-spin couplings. \\
This approach has been followed in several proposals for QS with molecular spin qubits. In Ref. \cite{SciRepNi}, the Authors considered a pair of Cr$_7$Ni qubits with an interposed Ni switch to show a proof-of-principle QS of the transverse field Ising model as presented above. $n=10$ Suzuki-Trotter steps were used to obtain a good approximation of the target system dynamics, covering the first non-trivial oscillations of the magnetisation. This yields a pulse sequence consisting of 10 two-qubit gates and 20 rotations which can be implemented in times significantly shorter than $T_2$, thus well reproducing the expected behaviour of the target observable. \\
Other spin models with non Ising couplings could benefit from a native two-qubit gate different from the controlled-phase shift. In particular, the $U_{XY}$ gate introduced in Eq. \eqref{eq:Uxy} directly implements an interaction of the form $s_{x1}s_{x2}+s_{y1}s_{y2}$ and can be easily complemented with single qubits rotations to implement other interactions, such as Heisenberg. This was shown in Ref. \cite{Chem}, where the dynamics resulting from the sequence of pulses implementing the target evolution was simulated including the effect of decoherence on a pair of Cr$_7$Ni rings linked via a Ru$_2$Co triangle. The latter shows a spin 1/2 ground state which could be oxidized to a diamagnetic state, thus effectively swithing on and off the coupling between the two spectroscopically equivalent Cr$_7$Ni qubits.\\
\begin{figure}[b]
    \centering
    \includegraphics[width=0.47\textwidth]{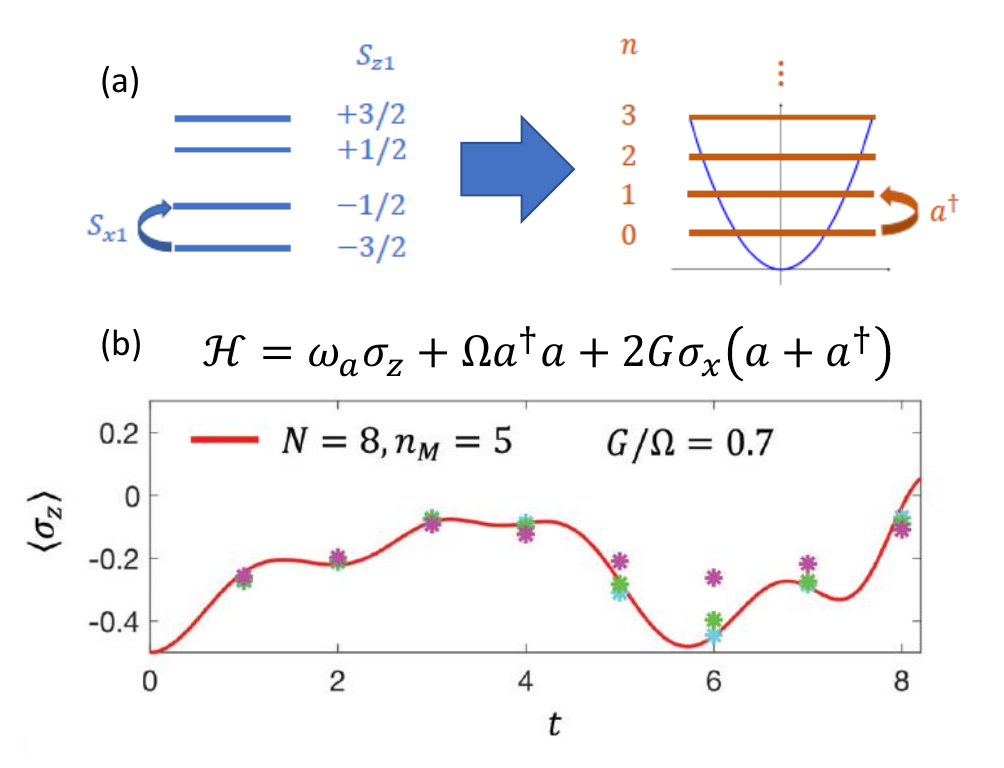}
    \caption{ Quantum simulation of the interaction between a spin and a photon in the strong coupling regime. (a) Mapping of the photon filed into the $2S+1$ states of a spin $S$ qudit. (b) Numerical simulation of the time evolution of $\sigma_z$, including decoherence and the full sequence of pulses needed to reproduce the dynamics of the target Hamiltonian (reported above). Cyan, green and magenta symbols represents simulations with $T_2=\infty, 10, 50 \, \mu$s, respectively. Reprinted with permission from \cite{TacchinoQudits}. }
    \label{fig:spinboson}
\end{figure}
\begin{figure*}[t!]
    \centering
    \includegraphics[width=0.9\textwidth]{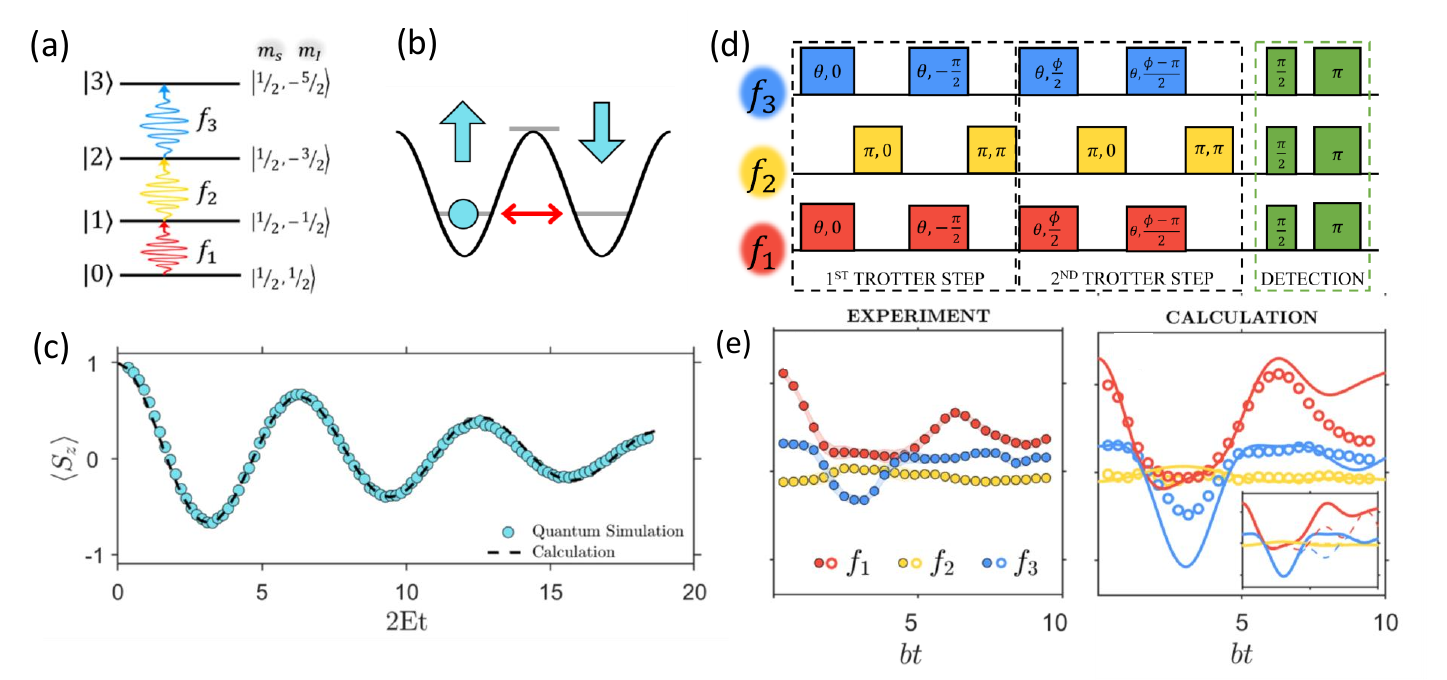}
    \caption{Proof-of-principle molecular spin quantum simulator based on $^{173}$Yb(trensal) nuclear spin qudits. (a) Energy levels exploited to encode the state of the target systems and corresponding energy gap, addressed by resonant radio-frequency pulses. (b-c) Quantum simulation of the tunnelling of the magnetisation for a spin $S=1$ prepared in a $m=1$ state and evolving under the effect of an axial ($D<0$) and rhombic anisotropy terms in the zero-field splitting Hamiltonian. The resulting magnetisation of the target system (extracted by measuring population differences between neighbouring energy levels) oscillates between opposite values, as expected (c). (d) Sequence of multi-frequency pulses used to implement the time evolution induced by the transverse field Ising model on two qubits, whose state is encoded into the 4 qudit states reported inpanel (a). The final detection step (green rectangles) corresponds to an Hahn echo sequence on each transition. (e) Resulting population differences between neighbouring energy levels on the hardware, in good agreement with calculations, from which different observables can be obtained. Adapted with permission from \cite{Chicco2023}.}
    \label{fig:simulator}
\end{figure*}
Although MNMs were first proposed as units of a qubit-based QS, their multi-level structure is particularly suitable to efficiently map intrinsically multi-level problems, such as those involving spins larger than 1/2, bosonic or fermionic degrees of freedom.
The QS of bosons is a prototypical example. Qubit-based mappings usually involve a large number of qubits (even exponential in the number of boson states \cite{DiPaolo2020}) or require complex entangling gates between different qubits in the register to simulate boson creation/annihilation \cite{Sawaya2020,Mathis2020}. A qudit approach provides a much simpler solution: the truncated boson space can be mapped to the $d$ levels of the qudit.\\
For instance, a spin $S>1/2$ subject to a magnetic field along its easy-magnetisation axis can be described by the spin Hamiltonian:
\begin{equation}
    H_S = g \mu_B B S_z + D S_z^2,
\end{equation}
which commutes with $S_z$ and hence is characterised by the same eigenstates $\ket{m}$, such that $S_z \ket{m} = m \ket{m}$.
Therefore, $2S+1$ boson states can be encoded into these levels and the creation and annihilation operators $a^\dagger$ and $a$ are mapped onto $S_+=S_x+iS_y$ and $S_-=S_x-iS_y$, respectively, as sketched in Fig. \ref{fig:spinboson}-(a). This mapping introduces an approximation related to truncating the boson field, but this is usually very small and it is also common to other encodings. Indeed, interesting purely quantum behaviors such as the strong and ultra-strong atom-photon coupling are very well reproduced even with a few bosons and hence $S=3/2$ or $S=5/2$ molecular systems could already be suitable for this goal \cite{TacchinoQudits}. Simulation results with different values of the spin qudit $T_2$ are reported in Fig. \ref{fig:spinboson}-(b), in good agreement with the expected behaviour (continuous line). \\

The quantum simulation of fermionic problems is also an outstanding problem in condensed matter Physics. The standard way to address this QS is via the Jordan-Wigner transformation between fermionic ($\left\{ c_j,c_k^\dagger \right\} = \delta_{jk}$) and spin operators:
\begin{subequations}
\begin{eqnarray}
&& c_j^\dagger c_j = \frac{1+\sigma_j^z}{2} \\
c_j^\dagger &=& e^{i \pi \sum_{k=1}^{j-1} c^\dagger_k c_k} \sigma_j^+ = \left(\prod_{k=1}^{j-1}  -\sigma_k^z \right) \sigma_j^+ \\
c_j &=& e^{-i \pi \sum_{k=1}^{j-1} c^\dagger_k c_k} \sigma_j^- = \left(\prod_{k=1}^{j-1}  -\sigma_k^z \right) \sigma_j^- .
    \label{eq:JordanWigner}
\end{eqnarray}
\end{subequations}
For multi-dimensional problems, this mapping introduces non-local many-body terms in the Hamiltonian $H$ to be simulated, whose decomposition into elementary operations on the hardware can be very complex and increasing with the size of the target problem. Qudits could facilitate the possibility to exploit encodings different from Jordan-Wigner, which remove the need of many-body gates by adding auxialiry levels to each computational unit \cite{PRA_JW_Laflamme,PRA_Fermion}.  \\

A proof-of-concept QS based on molecular spin qudits was recently demonstrated  \cite{Chicco2023}, using a nuclear spin 5/2 $^{173}$Yb(trensal) qudit  manipulated by a broad-band nuclear-magnetic resonance spectrometer [Fig. \ref{fig:simulator}-(a)]. Being an intrinsic multi-level system, it is suitable to target a multi-level problem, such as the dynamics of a spin $S>1/2$. In particular, the Authors considered a prototypical problem in molecular magnetism: the quantum tunnelling of the magnetisation which affects a spin $S$ subject to the following target Hamiltonian:
\begin{equation}
\mathcal{H} = D S_z^2 + E(S_x^2-S_y^2),
    \label{eq:tunneling}
\end{equation}
with a leading easy-axis axial anisotropy ($D<0$) and a small transverse term $E$. 
If the system is prepared in one of the two degenerate low energy states with $|M|=S$, it is expected to undergo oscillations between $M=-S$ and $M=S$ with a rate given by $E$, as sketched in Fig. \ref{fig:simulator}-(b).
A qubit-based simulation of this model would require $2S$ qubits prepared in the state of maximum magnetisation and manipulated in a symmetric way, in order never to leave the subspace with maximum total spin \cite{Santini2011,VO2}. In the qudit-based proof-of-concept demonstration, 3 levels of the $I=5/2$ nuclear qudit are exploited and the expected oscillations of the target magnetization are very nicely reproduced [see Fig. \ref{fig:simulator}-(c)]. \\
Moreover, the qudit states can be exploited to map the states of multiple spins 1/2, thus in fact reducing the number of error-prone two-body gates required in a given simulation. Indeed, these are usually the most difficult and error-prone operations to be implemented. In Ref. \cite{Chicco2023}, the transverse field Ising model discussed above is considered. The four-dimensional Hilbert space of the two spins is mapped onto four levels of the $I=5/2$ nuclear qudit and the time-evolution operator is decomposed into a sequence of pulses between neighboring ($\Delta m = \pm 1$) levels of the qudit [see Fig. \ref{fig:simulator}-(d)]. Finally, population differences between the eigenstates are detected by Hahn-echo sequences and evolution of some observables on the target system is reconstructed, again in good agreement with calculations [see Fig. \ref{fig:simulator}-(e)].

\section{Molecular Nanomagnets for Quantum Information Processing: potentialities and challenges}
\label{sec:advantages}
We conclude this introductory overview with a discussion to summarise the most important peculiarities of Molecular Nanomagnets which could represent important potentialities for their use as elementary building-blocks of a future quantum hardware. In parallel, we present the challenges which still need to be overcome for their actual employment. \\
First of all, MNMs are rather unique system for the degree of control on their magnetic properties which can be achieved at the synthetic level. This allows for a close collaboration between theoreticians and experimental chemists to design structures tailored for specific quantum applications. For instance, decoherence can be strongly suppressed by engineering both the couplings in the spin Hamiltonian \cite{Troiani2008,Chiesa2022} (as outlined in Sec. \ref{sec:deco}) and low-energy vibrations \cite{Wedge2012}, reaching very long values of $T_2$ \cite{Zadrozny2015}. \\
The pattern and the coupling between the magnetic ions can also be tuned to a large extent, fulfilling the requirements to implement, e.g., an effectively switchable interaction between logical qubits. 

Moreover, MNMs are natural multi-level spin systems to encode {\it qudits}, a potentially very important resource to improve the power of noisy devices (by reducing, for instance, the number of error-prone two-body gates in many algorithms) and also to embed quantum error correction within a single object. This is a remarkable difference from qubit-based platforms, in which the actual implementation of quantum error correction requires an enormous overhead of physical qubits and gates \cite{Mariantoni2012,JPCLqec,Mezzadri}. 
We stress that, at difference from other quantum systems which are also characterised by several energy levels, in MNMs coherence is in general not suppressed by increasing the number of states embedded in the qudit \cite{JPCLqec}. \\
Several methods based on microwave or radio-frequency electro-magnetic pulses allow for an accurate control of molecular spin qudits. Again, the design of pulse sequences can be combined with the engineering of the spin Hamiltonian to get the suitable pattern of allowed magnetic dipole transitions. This spin Hamiltonian design has enabled the first demonstration of a fault-tolerant protocol for quantum computing on MNMs \cite{Mezzadri}, where dephasing errors are suppressed even during the implementation of a universal set of gates. 

Spin Hamiltonian engineering, intrinsic multi-level structure, remarkable coherence properties, easy manipulation tools are powerful features which make MNMs a promising class of systems in the current rush for the physical implementation of QIP. However, they have still been little explored at the experimental level. The most important challenge to win is represented by reading out the state of a single molecular spin. Due to its tiny magnetic moment, this is not an easy task, but several routes are being investigated. The most promising one is probably based on reaching the strong coupling regime between individual spins and photons in superconducting resonators \cite{Carretta2021,Chiesa2023prappl}. Indeed, this architecture would also allow for scaling up the architecture in analogy to superconducting qubits. Combined with synthetic Chemistry methods, this approach could represent the asset for the success of this platform.

\section{Acknowledgements}
This work received financial support from European Union – NextGenerationEU, PNRR MUR project PE0000023-NQSTI.

\section{Biographies}

\paragraph*{Alessandro Chiesa}
Alessandro Chiesa got the PhD in Physics (2016) at the University of Parma. In 2017 he worked as postdoc at Jülich Forschungszentrum (Germany), with Angelo Della Riccia fellowship and then in the Spin-based Quatum Science Group at the University of Parma, where he is Research Associate since 2020. He teaches Quantum Computing at the Master Degree in Computer Sciences. His research interests span from the theoretical modelling of Molecular Nanomagnets to investigate fundamental quantum mechanical issues to the development of schemes for Quantum Information Processing (Quantum Simulation, Quantum Error Correction, Quantum Computation) with Molecular Spin Qudits and other physical systems, in close connection with experiments.
\\
\paragraph*{Emilio Macaluso}
Emilio Macaluso studied Physics at the University of Parma, where he got his PhD in Physics in 2021. During his doctoral studies in the Spin-based Quantum Science group, he concentrated on investigating molecular nanomagnets in the field of information technology, employing a combination of experimental and theoretical techniques. Following the completion of his PhD, he continued working within the Spin-based Quantum Science Group, where his main research focus has been chirality-induced spin selectivity in electron transfer phenomena and its potential applications in quantum technologies. Since 2023, he is a Research Associate within the same group, working on the broader field of molecular spin qub(d)its in the field of Quantum Information Processing.
\\
\paragraph*{Stefano Carretta}
Stefano Carretta obtained his PhD in Physics in 2005 and is Full Professor of Theoretical Physics of Matter at the University of Parma since 2019. His research activity encompasses several fields, from magnetism to quantum technologies. In particular, he contributed to put forward most of the protocols for quantum information processing with molecular nanomagnets. He is coauthor of more than 160 research papers and has been scientific responsible in several research projects focused on quantum technologies.
He received the "Le Scienze" and President of the Italian Republic medals in 2006 and the Olivier Kahn International Award in 2011 for his contributions to molecular magnetism.

\bibliography{bibliography}

\end{document}